\documentclass[useAMS,usenatbib]{mn2e}
\usepackage{graphicx}
\pdfoutput=1 

\def\kms{{km~s}$^{-1}$}
\def\nai{Na\,{\sc I}}
\def\cai{Ca\,{\sc I}}
\def\fei{$<$Fe\,{\sc I}$>$}
\def\feia{Fe\,{\sc I}\,A}
\def\feib{Fe\,{\sc I}\,B}
\def\mgi{Mg\,{\sc I}}
\def\caviii{[Ca\,{\sc VIII}]}

\title[The SMBH of Fornax A]{The supermassive black hole of Fornax\,A\thanks{Based on observations at the European Southern Observatory VLT (076.B-0457(A)) and archival ESO La Silla (66.C-0310(A)) and NASA/ESA {\it Hubble Space Telescope} data (GO Proposal 7458), obtained from the ESO/ST-ECF Science Archive Facility}}
\author[N. Nowak et al.]{N. Nowak$^{1,2}$\thanks{E-mail:
nnowak@mpe.mpg.de}, R. P. Saglia$^{1,2}$, J. Thomas$^{1,2}$, R. Bender$^{1,2}$, R. I. Davies$^{1}$, K. Gebhardt$^{3}$\\
$^{1}$Max--Planck--Institut f\"{u}r extraterrestrische Physik, Giessenbachstrasse, 85748 Garching, Germany\\
$^{2}$Universit\"{a}tssternwarte, Scheinerstrasse 1, 81679 M\"{u}nchen, Germany\\
$^{3}$Astronomy Department, University of Texas, Austin, TX~78723}
\begin{document}

\date{Received 2008 August 22; in original form 2008 July 19}

\pagerange{\pageref{firstpage}--\pageref{lastpage}} \pubyear{2008}

\maketitle

\label{firstpage}

\begin{abstract}
The radio galaxy Fornax\,A (NGC\,1316) is a prominent merger remnant
in the outskirts of the Fornax cluster. Its giant radio lobes suggest
the presence of a powerful AGN, and thus a central supermassive black
hole (SMBH). Fornax\,A now seems to be in a transition state between
active black hole growth and quiescence, as indicated by the strongly
declined activity of the nucleus. Studying objects in this
evolutionary phase is particularly important in order to understand
the link between bulge formation and black hole growth, which is
manifested in the $M_\bullet$-$\sigma$ relation between black hole
mass and bulge velocity dispersion. So far a measurement of the SMBH
mass has not been possible in Fornax\,A, as it is enshrouded in dust
which makes optical measurements impossible. We present
high-resolution adaptive optics assisted integral-field data of
Fornax\,A, taken with SINFONI at the Very Large Telescope in the $K$
band, where the influence of dust is negligible. The achieved spatial
resolution is $0.085$~arcsec, which is about a fifth of the diameter
of the expected sphere of influence of the black hole.  The stellar
kinematics was measured using the region around the CO bandheads at
$2.3$~$\umu$m. Fornax\,A does not rotate inside the inner
$\sim3$~arcsec. The velocity dispersion increases towards the
centre. The weak AGN emission affects the stellar kinematics in the
inner $\sim0.06$~arcsec only. Beyond this radius, the stellar
kinematics appears relaxed in the central regions. We use axisymmetric
orbit models to determine the mass of the SMBH in the centre of
Fornax\,A. The three-dimensional nature of our data provides the
possibility to directly test the consistency of the data with
axisymmetry by modelling each of the four quadrants
separately. According to our dynamical models, consistent SMBH masses
$M_\bullet$ and dynamical \emph{Ks} band mass-to-light ratios
$\Upsilon$ are obtained for all quadrants, with $\langle
M_{\bullet}\rangle=1.3\times10^{8}$~M$_{\odot}$
($\mathrm{rms}(M_{\bullet})=0.4\times10^{8}$~M$_{\odot}$) and
$\langle\Upsilon\rangle=0.68$ ($\mathrm{rms}(\Upsilon)=0.03$),
confirming the assumption of axisymmetry. For the folded and averaged
data we find $M_{\bullet}=1.5_{-0.8}^{+0.75}\times10^8$~M$_{\odot}$
and $\Upsilon=0.65^{+0.075}_{-0.05}$ ($3\sigma$ errors). Thus the
black-hole mass of Fornax\,A is consistent within the error with the
\citet{Tremaine-02} $M_{\bullet}$-$\sigma$ relation, but is a factor
$\sim4$ smaller than expected from its bulge mass and the
\citet{MarconiHunt-03} relation.

\end{abstract}


\section{Introduction}

Studies of the dynamics of stars and gas in the centres of nearby
galaxies with a massive bulge component have established the presence
of supermassive black holes (SMBHs) in the $10^6$--$10^9$~M$_\odot$
range. The mass of the central SMBH is tightly correlated with the
bulge mass or luminosity (\textrm{e.g.} \citealt{MarconiHunt-03}) and
with the bulge velocity dispersion $\sigma$
\citep{Gebhardt-00a,Ferrarese-00}. These correlations suggest that
bulge evolution and black hole growth are closely linked.  Indeed
there is increasing theoretical evidence that galaxy merging (or other
processes that lead to gas inflow, like secular evolution), nuclear
activity and feedback are all somehow linked to bulge and SMBH
evolution (\textrm{e.g.}
\citealt{diMatteo-05,Johansson-08,Younger-08,Hopkins-08a}).  To
observationally constrain present theories of bulge and black hole
evolution, detailed studies of AGN and merger remnants in different
evolutionary stages are essential. NGC\,5128 (Cen\,A) is presently the
only galaxy with a powerful AGN that underwent a recent major merger
which has a measured black-hole mass
\citep{Silge-05,Marconi-06,Neumayer-07}.

A galaxy very similar to Cen\,A is NGC\,1316 (Fornax\,A).  Fornax\,A
is a giant elliptical galaxy located in the outskirts of the Fornax
galaxy cluster. It is one of the brightest radio galaxies in the sky
with giant double radio lobes and $\mathcal{S}$-shaped
nuclear radio jets \citep{Geldzahler-84}. The peculiar morphology with
numerous tidal tails, shells and loops
\citep{Schweizer-80,Schweizer-81} suggests that a major merger
happened about $3$~Gyr ago \citep{Goudfrooij-01}, followed by some
minor mergers \citep{Mackie-98}. Its nucleus however is surprisingly
faint in X-rays, which can only be explained if the nucleus became
dormant during the last $0.1$~Gyr \citep{Iyomoto-98}. This makes it
an ideal target for a dynamical black-hole mass measurement, as its
stellar absorption lines are probably not or just marginally diluted
by non-stellar emission from the AGN.

Fornax\,A is classified as an intermediate form between core and power
law galaxy in \citet{Lauer-07}. The surface brightness approximately
follows an $r^{1/4}$-law \citep{Caon-94}. Large amounts of dust are
present also in the inner few arcseconds, which affects optical
spectroscopy. We therefore measure the two-dimensional stellar
kinematics of Fornax\,A with the near-infrared integral-field
spectrograph SINFONI at the Very Large Telescope (VLT) in the $K$
band, where the dust obscuration is only about $7$~per~cent of the
obscuration in the optical. We use this data to analyse the
stellar kinematics in a way similar to \citet{Nowak-07}.

Throughout this paper we adopt a distance to Fornax\,A of $18.6$~Mpc
based on \emph{HST} measurements of Cepheid variables in NGC\,1365
\citep{Madore-99}. At this distance, $1$~arcsec corresponds to
$90$~pc. With a velocity dispersion of $226$~\kms\ (mean $\sigma$
measured in an $8$~arcsec aperture in \S 3.4) the estimated sphere of
influence has a diameter of $0.46$~arcsec and we would expect a SMBH
mass of $2.2\times10^8$~M$_\odot$ \citep{Tremaine-02}. This is large
enough to be resolved easily from the ground with adaptive optics.

This paper is organized as follows: In \S 2 we present the data and
the data reduction. The stellar kinematics of Fornax\,A is described
in \S 3, the measurement of the near-infrared line strength indices in \S
4. The stellar dynamical modelling procedure and the results for the
SMBH mass are presented in \S 5, and \S 6 summarises and discusses the
results.

\section{Data \& Data Reduction}

\subsection{SINFONI data}

Fornax\,A was observed between 2005 October 10 and 12 as part of
guaranteed time observations with SINFONI
\citep{Eisenhauer-sinfoni,Bonnet-sinfoni}, an adaptive-optics assisted
integral-field spectrograph at the VLT UT4.  The nucleus of Fornax\,A 
served as guide star for the AO correction. We used the $K$ band
grating ($1.95-2.45$~$\umu$m) and, depending on the seeing conditions,
the intermediate size field of view of $3\times3$~arcsec
($0.05\times0.1$~arcsec~spaxel$^{-1}$, referred to as ``100mas
scale'' in the following) or the high resolution mode with a
$0.8\times0.8$~arcsec field of view
($0.0125\times0.025$~arcsec~spaxel$^{-1}$, ``25mas scale''). The
total on-source exposure time was $130$~min in the highest resolution
mode and 70~min using the intermediate plate scale, consisting of
$10$~min exposures taken in series of ``object--sky--object'' (O--S--O)
cycles, dithered by up to $0.2$~arcsec. In addition three O--S--O cycles
of $5$~min exposure time each were taken using the low resolution mode
($8\times8$~arcsec field of view, ``250mas scale'').

\begin{figure*}
\begin{minipage}{168mm}
\centering
\includegraphics[width=\linewidth,keepaspectratio]{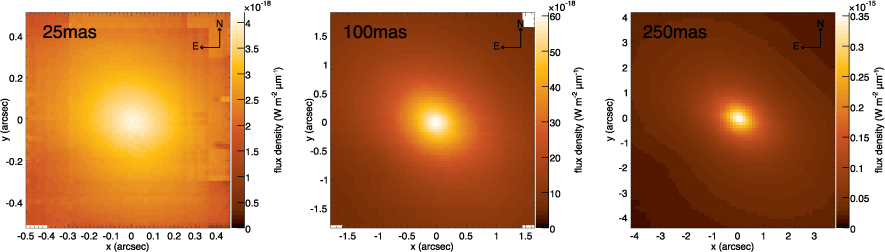}
  \caption{SINFONI images of Fornax\,A in the 25mas (left), 100mas (middle) and 250mas (right) scale.}\label{fig:1}
\end{minipage}
\end{figure*}

The SINFONI data reduction package {\sc SPRED}
\citep{Schreiber-spred,Abuter-spred} was used to reduce the data. It
includes all common reduction steps necessary for near-infrared
(near-IR) data plus routines to reconstruct the three-dimensional
datacubes. The data were skysubtracted, flatfielded, corrected for bad
pixels and for distortion and then wavelength calibrated using a Ne/Ar
lamp frame. The wavelength calibration was corrected using night-sky
lines if necessary. Then the datacubes were reconstructed and
corrected for atmospheric absorption using B stars that do not have
strong spectral features in the region of the CO bandheads. As a final
step all datacubes were averaged together to produce the final
datacube. The data of the telluric stars were reduced likewise.
The flux calibration of the 250mas data was performed by comparison to
the photometrically calibrated NTT/SOFI {\emph Ks} band imaging (see
\S 2.3 and Fig. \ref{fig:3}), that gives a {\emph Ks} magnitude of
$8.44$ in an $8$~arcsec aperture. The 100mas image was calibrated with
respect to the 250mas image and the 25mas image with respect to the
100mas image. Fig. \ref{fig:1} shows the resulting image of Fornax\,A
in all three platescales, collapsed along the wavelength direction
within the SOFI {\emph Ks} band region.

The spatial resolution, \textrm{i.e.} the FWHM of the point-spread function
(PSF), is difficult to derive when dealing with high-resolution AO
observations of diffuse objects. 
Deriving it from the science data itself can work quite well, if there
are either stars in the field of view or the target is an AGN or
starburst galaxy with certain spectral properties (see \textrm{e.g.}
\citealt{Mueller-06,Davies-06,Davies-07b}). None of this applies to
Fornax\,A. Therefore we derived the PSF by regularly taking an
exposure of a nearby star with approximately the same $R$ band
magnitude and $B-R$ colour as the central $3$~arcsec of Fornax\,A.
This procedure is usually recommended to observers, but apart from
being very time-consuming it can also be unreliable under certain
conditions \citep{Davies-04}.  The atmosphere may vary strongly with
time, resulting in a measured PSF that is different from the real
one. In addition the response of the wavefront sensor to the PSF star
is, to some degree, different from the response to the AO guide star
(the nucleus in this case), because the nucleus is extended and
therefore the flux distribution on the wavefront sensor is
different. We used this method as it was the only available option to
measure the PSF.

\begin{figure}
\centering
\includegraphics[height=.9\textheight,keepaspectratio]{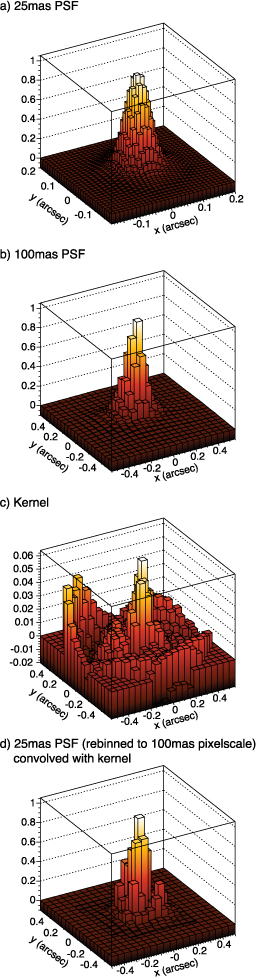}
  \caption{a) 25mas PSF derived by observing a star of about the same
  magnitude and colour as the AO guide star (\textrm{i.e.} the nucleus of
  Fornax\,A) during the observations. b) Same as (a) but for the
  100mas scale. c) Kernel that transforms the 25mas image of Fornax\,A
  to the 100mas image. d) 25mas PSF from (a) rebinned to the 100mas
  spaxel size and convolved with the kernel from (c).}\label{fig:2}
\end{figure}

The high-resolution (25mas) data were taken under good and relatively
stable ambient conditions with a near-IR seeing around $0.6$~arcsec. The
two-dimensional PSF (\textrm{i.e.} the image of the PSF reference star) is
shown in Fig.  \ref{fig:2}a. The spatial resolution is
$\mathrm{FWHM}\approx0.085$~arcsec. A Strehl ratio of $\approx45$~per~cent was
reached. The reliability of this PSF is verified by a comparison of the
25mas SINFONI luminosity profile with an \emph{HST}/NICMOS F160W (camera
NIC2, sampled at $0.075$~arcsec~px$^{-1}$) luminosity profile. They
agree well without further broadening of the NICMOS data
(see below and Fig. \ref{fig:4}).

During the hours of degraded seeing (around $\approx0.9$~arcsec on
average in the near-IR) on October 12 we observed Fornax\,A using the
intermediate plate scale (100mas). The spatial resolution inferred
from the PSF is $\approx0.16$~arcsec and the achieved Strehl ratio was
$\approx18$~per~cent.  The shape of the PSF (Fig. \ref{fig:2}b) is rather
asymmetric and may not represent the true PSF due to the changeable
weather conditions. The seeing and the coherence time drastically
improved during the observations with the 100mas scale, so we switched to
the 25mas scale for approximately one hour and only when the seeing
and the coherence time then degraded again to similar values as before
we could observe the 100mas PSF.
In order to get a better estimate of the real PSF we computed the
kernel that transforms the 25mas image of Fornax\,A (binned $4\times4$
to match the 100mas spaxel size) into the 100mas image using the
program of \citet{Goessl-02}. The 25mas PSF was then rebinned to the
100mas spaxel size and convolved with this kernel. Kernel and
resulting PSF are shown in Fig. \ref{fig:2}c and \ref{fig:2}d. The
shape of the kernel and hence the convolved PSF are quite noisy.  The
reason is the low S/N of the 25mas image compared to the 100mas image
(\textrm{cf.} Fig. \ref{fig:1}). In order to find the closest match between
the two images the kernel tries to incorporate the noise, too. Thus the
resulting PSF has a S/N more similar to the 25mas image than to the
image of a bright PSF star. Nevertheless the convolved PSF has about
the same FWHM as the measured 100mas PSF star.

The PSF is an integral part of the dynamical modelling and it is
therefore important to know its shape as accurate as possible. However,
so far no studies have been performed that analyse systematically the
dependence of the resulting $M_{\bullet}$ and mass-to-light ratio on the PSF shape. Such a
study is beyond the scope of this paper, nevertheless we will do the
modelling, as far as 100mas data are included, twice, using one time
the measured PSF and the other time the convolved PSF.

\subsection{Longslit data}

Kinematics derived from longslit spectra are useful to constrain the
orbital structure at large radii, outside the small SINFONI field of
view.  Major-axis longslit data of Fornax\,A are available from
different authors
\citep{Longhetti-98,Arnaboldi-98,Bedregal-06,Saglia-02}. \citet{Longhetti-98}
and \citet{Arnaboldi-98} measured only $v$ and $\sigma$ from optical
absorption lines (Mg$b$) and from the CaT region in the near-IR, they
did not measure the higher-order Gauss--Hermite coefficients $h_3$ and
$h_4$ \citep{Gerhard-93,Marel-93} which quantify asymmetric and
symmetric deviations from a Gaussian velocity
profile. \citet{Bedregal-06} measured all four parameters from optical
spectra, whereas \citet{Saglia-02} used the CaT region. The velocities
of all measurements are consistent, taking into account the different
seeing values. The velocity dispersions in the central $\sim5$~arcsec
however range from $\approx220$~\kms\ to $\approx260$~\kms. The
reasons for that can be diverse. The authors used different
correlation techniques, different spectral lines and slightly
different position angles. Fornax\,A contains dust, which might alter
the kinematics measured from optical absorption
lines. \citet{Silge-03} found that optical dispersions tend to be
larger than those measured from the CO bandheads, which could be an
effect of dust. The largest dispersions are indeed those measured from
optical spectra. We therefore used the kinematics of \cite{Saglia-02},
as they measured all four parameters from the CaT line region. Their
measurements agree best with our SINFONI measurements. They observed
Fornax\,A in October 2001 at the Siding Spring $2.3$m telescope using
a $6.7$~arcmin$\times4$~arcsec longslit and an exposure time of
$60$~min and determined the kinematics with the Fourier Correlation
Quotient method \citep{Bender-94}. As they assume a position angle of
$58\degr$ we adopt this value in the following analysis. The major
axis profiles are shown in Fig. \ref{fig:17} and in
\citet{Beletsky-08}.

\subsection{Imaging}

\begin{figure}
\centering
\includegraphics[width=.95\linewidth,keepaspectratio]{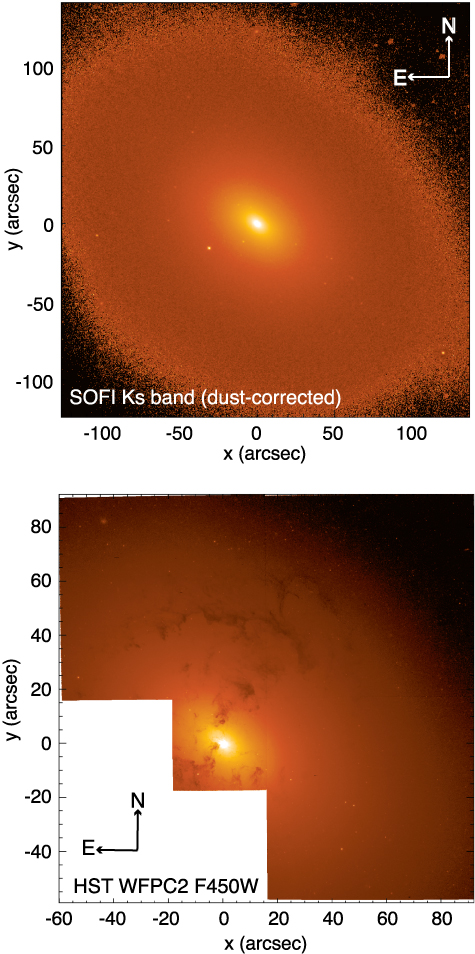}
  \caption{The top panel shows the dust-corrected SOFI \emph{Ks} band
  image of Fornax\,A (courtesy Y. Beletsky). The dust-corrected light
  distribution appears very smooth and regular and the isophotes are
  slightly boxy. For comparison the \emph{HST} WFPC2 F450W image
  (NASA/ESA \emph{HST} data (GO Proposal 5990), obtained from the
  ESO/ST-ECF Science Archive Facility), which highlights the strong
  dust lanes, is presented in the bottom
  panel. }\label{fig:3}
\end{figure}

\begin{figure}
\centering
\includegraphics[width=.95\linewidth,keepaspectratio]{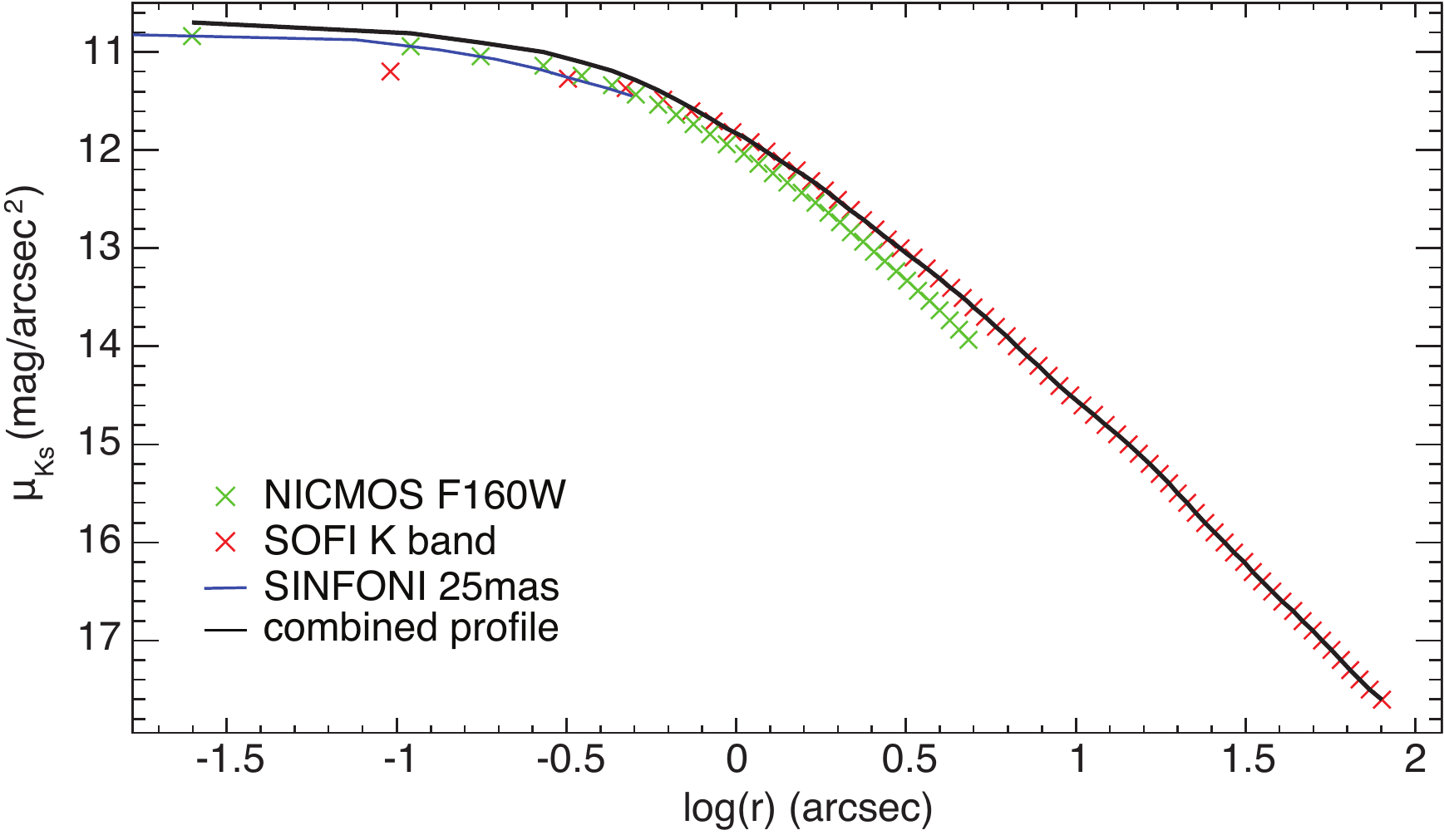}
  \caption{Surface brightness profiles of the \emph{HST} NICMOS
  dust-corrected $H$ band image (green crosses), the SOFI dust
  corrected \emph{Ks} band image (red crosses) and the combination of the
  two surface brightness profiles (black line). The SINFONI 25mas $K$
  band image (scaled to match the \emph{HST} NICMOS profile) is overplotted
  in blue.}\label{fig:4}
\end{figure}

To measure the SMBH mass in Fornax\,A, it is essential to determine
the gravitational potential made up by the stellar component using
photometric measurements with sufficient spatial resolution and radial
extent. Therefore we combine high-resolution \emph{HST} NICMOS F160W
and ground-based wide-field \emph{Ks} band imaging. The low-resolution
\emph{Ks} band image was taken with the near-IR imager/spectrometer
SOFI on the $3.5$m NTT telescope on La Silla, Chile. It has a field of
view of $4.9\times4.9$~arcmin with $0.29$~arcsec~px$^{-1}$ and a
seeing of $\sim0.7$~arcsec. It was dust-corrected with the method
described in Appendix A using a $J$ band SOFI image. The
dust-corrected image is shown in Fig. \ref{fig:3}. Note that the
original SOFI image and the residuals are shown in
\citet{Beletsky-08}. We applied the same method to the \emph{HST}
NICMOS F160W image, which we corrected using a NICMOS F110W image. The
isophotal profiles of the dust-corrected images were obtained
following \citet{Bender-87}. The light distribution appears very
smooth and regular and the isophotes are slightly boxy. The surface
brightness profiles were then combined using the method of
\citet{Corsini-08}, matching the profiles in the region between
$1.5$~arcsec and $5.0$~arcsec and shifting the NICMOS profile by the
computed amount ($\mu_{\mathrm{SOFI}}-\mu_{\mathrm{\emph{HST}}}$) on
to the \emph{Ks} band SOFI profile.  The original and the combined
profiles are shown in Fig. \ref{fig:4}.

The combined photometry extends out to $80$~arcsec and was deprojected
for an inclination $i=90\degr$ using the program of
\citet{Magorrian-99} under the assumption that the galaxy is
axisymmetric. No shape penalty was applied. The stellar mass density
then can be modelled via $\rho_{*}=\Upsilon\nu$, where $\nu$ is the
luminosity density obtained from the deprojection. The \emph{Ks} band
mass-to-light ratio $\Upsilon$ is assumed to be constant with
radius. This assumption holds approximately in the central part of the
galaxy for which we have kinematic data and where dark matter does not
play a major role.

\section{Stellar kinematics}

The stellar kinematic information are extracted using the maximum
penalised likelihood (MPL) technique of \citet{Gebhardt-00c}, which
obtains non-parametric line-of-sight velocity distributions (LOSVDs)
as follows: an initial binned velocity profile is convolved with a
linear combination of template spectra and the residuals of the
resulting spectrum to the observed galaxy spectrum are calculated. The
velocity profile is then changed successively and the weights of the
templates are adjusted in order to optimise the fit to the observed
spectrum by minimizing the function
$\chi^{2}_{\mathrm{P}}=\chi^{2}+\alpha\mathcal{P}$, where $\alpha$ is
the smoothing parameter that determines the level of regularisation,
and the penalty function $\mathcal{P}$ is the integral of the square
of the second derivative of the LOSVD. We fitted only the first two
bandheads $^{12}$CO($2-0$) and $^{12}$CO($3-1$), \textrm{i.e.} the spectral
range between $2.275$~$\umu$m and $2.349$~$\umu$m rest frame wavelength. The
higher-order bandheads are strongly disturbed by residual atmospheric
features. At wavelengths $\lambda<2.29~\umu$m most absorption lines
are weak such that extremely high S/N spectra would be needed. The
strongest absorption line in that regime is \nai, but as its strength
is much higher in early-type galaxies than in single stars due to
enhanced silicon \citep{Silva-08} it cannot be fitted by the template
spectra.

\subsection{Initial parameters}
Before being able to obtain the final LOSVDs it is necessary to
quantify possible systematic offsets and to constrain the best initial
values for the smoothing parameter and the width of the LOSVD bins for
the given dataset (wavelength range, $\sigma$, $h_{3}$, $h_{4}$, S/N,
spectral resolution). Especially the selection of the smoothing
parameter $\alpha$ is a crucial step \citep{Merritt-97,Joseph-01}. If
$\alpha$ is chosen too high, the LOSVD is biased toward a flatter shape
and if it is too small the LOSVD is too noisy. In order to find the
best initial values for the Fornax\,A data Monte Carlo simulations on
a large set of model galaxy spectra have been performed. These are
described in detail in Appendix B.

Based on the simulations we conclude that with MPL reliable LOSVDs can
be obtained from the first two CO bandheads of the SINFONI data of
Fornax\,A when the chosen $\alpha$ is between $1$ and $10$ and the
$\mathrm{S/N}\ga30$.  The S/N of our data is very high (on average
$\sim70$ for the 25mas and the 250mas data, and $\sim140$ for the
100mas data). We are using $\alpha=8$ for the 25mas and 250mas data
and $\alpha=6$ for the 100mas data.

\subsection{Kinematic template stars}
We built up a small library of late-type (K and M) kinematic template
stars which were observed with SINFONI in the $K$ band during
commissioning and GTO observations between 2004 and 2006. In total
we have nine kinematic templates for the 25mas scale (spectral resolving
power $R\approx5000$), twelve templates for the 100mas scale
($R\approx4500$) and ten templates for the 250mas scale
($R\approx4000$). As shown in \citet{Silge-03} a correct value of the
velocity dispersion $\sigma$ can only be obtained from fits to the CO
bandheads, if the template has about the same intrinsic CO equivalent
width (EW) as the galaxy. The CO EWs of the Fornax\,A spectra were
measured using the definition and the velocity dispersion correction
of \citet{Silge-03}. The resulting values are between
$13.0\mathrm{\AA}$ and $14.5\mathrm{\AA}$. 

After excluding all stellar templates with EWs far below or above the
measured range, five templates remained for the 25mas scale, five for
the 100mas scale and four for the 250mas scale. As a cross-check and
to avoid template mismatching the kinematics was extracted first with
single templates. The results agreed, so no other template star had to
be excluded from the sample. Tab. \ref{tab:templates} shows the
used kinematic template stars with the according spectral types and CO
EWs. For a better comparison with the line strength indices discussed
in \S 4 the CO EWs were also measured using the definition of
\citet{Silva-08}.

\begin{table}
 \centering 
 \begin{minipage}{80mm}
%  \caption{Near-IR line strength indices of the stellar kinematic template stars used for the extraction of the kinematics.}\label{tab:templates}
%  \begin{tabular}{llccccccr}
%  \hline
%   Name     & spectral type & CO\footnote{CO equivalent width as defined by \citet{Silge-03}} & CO\footnote{CO equivalent width as defined by \citet{Silva-08}} & \nai\ & \cai\ & \feia\ & \feib\ & scale\\
% \hline
% HD12642    & K5 III    & 12.7  & 17.09 & 3.01 & 2.23 & 1.44 & 0.83 &  25mas  \\  
% HD163755   & K5/M0 III & 15.6  & 21.27 & 3.98 & 2.21 & 1.89 & 1.07 &  25mas  \\
% HD179323   & K2 III    & 11.1  & 14.08 & 1.80 & 1.31 & 0.90 & 0.68 &  25mas  \\ 
% HD198357   & K3 III    & 10.8  & 13.84 & 2.19 & 1.80 & 1.15 & 0.81 &  25mas  \\
% HD75022    & K2/3 III  & 12.0  & 15.06 & 2.12 & 2.24 & 0.90 & 0.42 &  25mas, 100mas  \\
% HD141665   & M5 III    & 15.8  & 22.26 & 3.96 & 3.24 & 2.22 & 1.05 & 100mas, 250mas  \\
% HD181109   & K5/M0 III & 12.9  & 17.79 & 2.64 & 2.17 & 1.07 & 0.77 & 100mas, 250mas  \\
% HD201901   & K3 III    & 11.4  & 14.43 & 2.24 & 2.58 & 0.44 & 0.42 & 100mas, 250mas  \\
% SA112-0595 & M0 III    & 12.8  & 17.61 & 2.15 & 1.48 & 1.00 & 0.56 & 100mas, 250mas  \\ %from Cohen et al 2003,  AJ 125, 2645-2663
  \caption{CO equivalents widths of the stellar kinematic template stars.}\label{tab:templates}
  \begin{tabular}{llccr}
  \hline
   Name     & spectral type & CO\footnote{Definition of \citet{Silge-03}}  & CO\footnote{Definition of \citet{Silva-08}} & scale\\
 \hline
 HD12642    & K5 III    & 12.7  & 17.09 &   25  \\ 
 HD163755   & K5/M0 III & 15.6  & 21.27 &   25  \\
 HD179323   & K2 III    & 11.1  & 14.08 &   25  \\ 
 HD198357   & K3 III    & 10.8  & 13.84 &   25  \\
 HD75022    & K2/3 III  & 12.0  & 15.06 &   25, 100  \\ 
 HD141665   & M5 III    & 15.8  & 22.26 &  100, 250  \\
 HD181109   & K5/M0 III & 12.9  & 17.79 &  100, 250  \\
 HD201901   & K3 III    & 11.4  & 14.43 &  100, 250  \\
 SA112-0595\footnote{N. Neumayer, priv. comm.} & M0 III    & 12.8  & 17.61 &  100, 250  \\ 
\hline
\end{tabular}

\medskip
Note. The CO equivalent widths are given in units of
$\mathrm{\AA}$. The spectral types are taken from \citet{Wright-03}
where available.  In the last column the SINFONI platescales, in which
the stars have been observed, are given.
\end{minipage}
\end{table}

\subsection{Error estimation}
The uncertainties on the velocity profiles are estimated using Monte
Carlo simulations \citep{Gebhardt-00c}. A galaxy spectrum is created
by convolving the template spectrum with the measured LOSVD. Then
$100$ realisations of that initial galaxy spectrum are created by
adding appropriate Gaussian noise. The LOSVDs of each realisation are
determined and used to specify the confidence intervals.

\subsection{The kinematics of Fornax\,A}
\begin{figure}
\centering
\includegraphics[width=\linewidth,keepaspectratio]{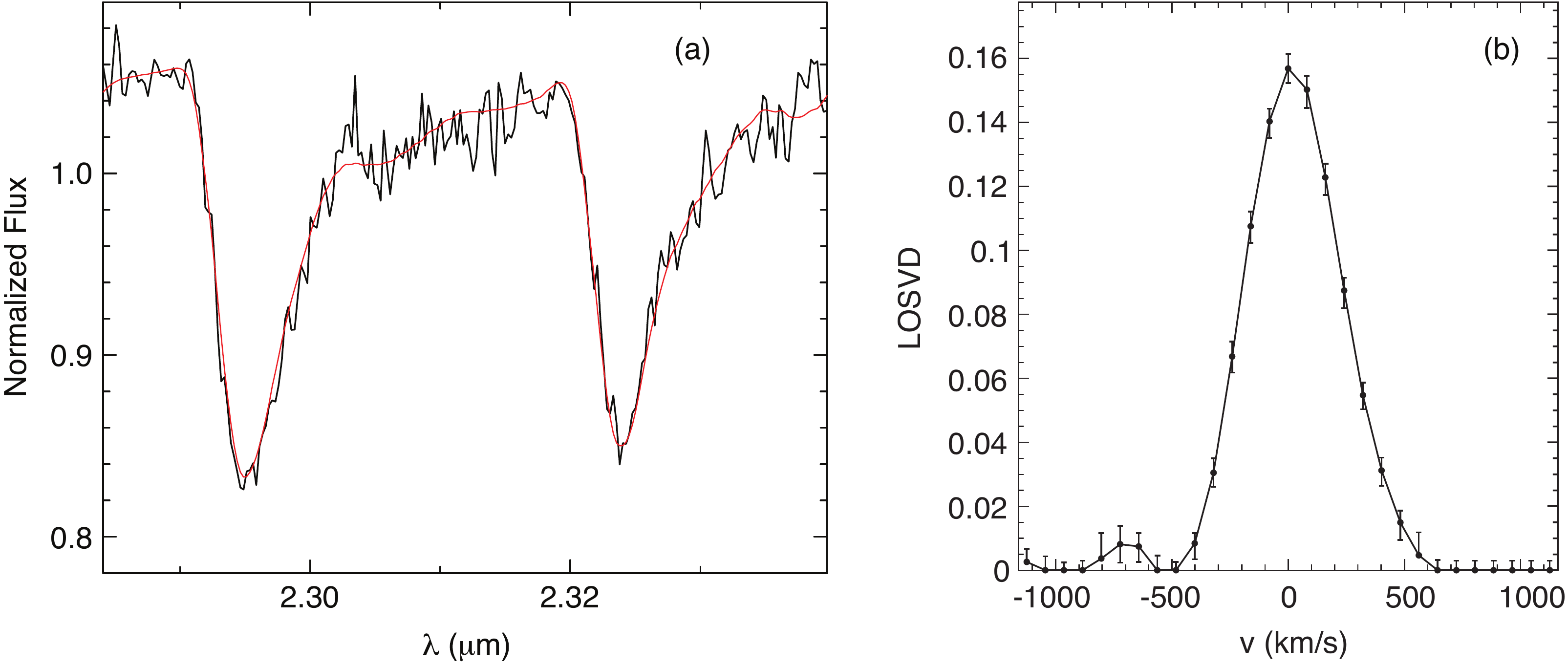}
  \caption{a) Example fit to a 25mas spectrum and b) the according LOSVD with error bars.}\label{fig:5}
\end{figure}

The SINFONI data are binned using a similar binning scheme as in
\cite{Nowak-07} with five angular bins per quadrant and a number of
radial bins (seven for the 25mas scale, ten for the 100mas scale and
13 for the 250mas scale). The centres of the angular bins are at
latitudes $\vartheta=5.8^{\circ}$, $17.6^{\circ}$, $30.2^{\circ}$,
$45.0^{\circ}$ and $71.6^{\circ}$. The stellar kinematics then was
extracted for all scales using the appropriate templates and $\alpha$.
Fig. \ref{fig:5} shows as an example the fit to a 25mas spectrum
and the according LOSVD.

\begin{figure*}
\begin{minipage}{168mm}
\centering
\includegraphics[width=22cm,keepaspectratio,angle=90]{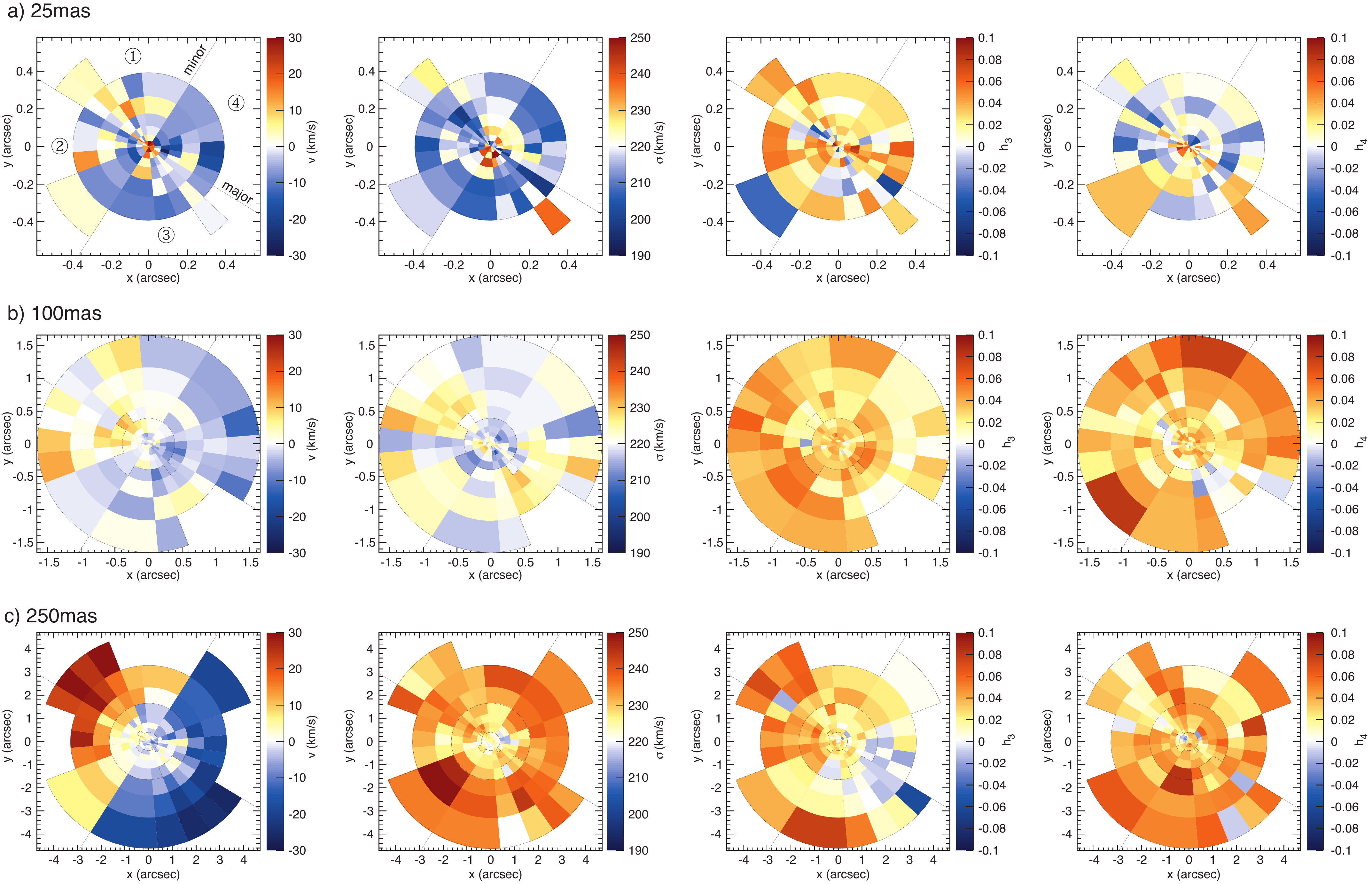}
  \caption{a) $v$, $\sigma$, $h_{3}$ and $h_{4}$ of Fornax\,A in the
  central $0.8$~arcsec (25mas scale). The major and
  the minor axis as well as the numeration of the different quadrants
  are indicated in the velocity field (left). b) shows the same
  as (a) but for the 100mas scale. The
  25mas field of view is marked for a better orientation. c) shows the
  same as (a) but for the 250mas scale. Both the 25mas and the 100mas
  field of view are marked.}\label{fig:6}
\end{minipage}
\end{figure*}

For the dynamical modelling we use these LOSVDs and no parametrized
moments, but for illustration purposes we fitted the LOSVDs with
Gauss--Hermite polynomials and we show the 2D fields of $v$, 
$\sigma$ and the Gauss--Hermite coefficients $h_{3}$ and $h_{4}$ in
Fig. \ref{fig:6} for all platescales.
The amplitudes of the errors are on
average about $8$~\kms\ ($v$ and $\sigma$) or $0.025$ ($h_{3}$ and
$h_{4}$) for the 25mas data, $4$~\kms\ or $0.015$ for
the 100mas data and $9.5$~\kms\ or $0.033$ for the 250mas data.

On large scales (250mas, Fig. \ref{fig:6}c)
the galaxy is clearly rotating and the mean velocity dispersion is
high ($\sigma\approx226$~\kms), while in the central $\sim3$~arcsec
(100mas, Fig. \ref{fig:6}b) no clear rotation is visible and
the dispersion is lower ($\sigma\approx221$~\kms). The 25mas velocity
field (Fig. \ref{fig:6}a) is irregular and with a peak in the
central bins. The velocity dispersion also rises towards the centre,
but the maximum does not coincide with the photometric centre. It is
located $\approx0.05-0.1$~arcsec south of the centre. North of
the centre there is a second but less strong maximum. The two maxima
are separated by a narrow, elongated $\sigma$-minimum. Outside the
central region the mean dispersion of the 25mas field is lower than in
the larger fields ($\sigma\approx218$~\kms). $h_{3}$ and $h_{4}$ are
on average small and positive.

\begin{figure*}
\begin{minipage}{168mm}
\centering
\includegraphics[width=\linewidth,keepaspectratio]{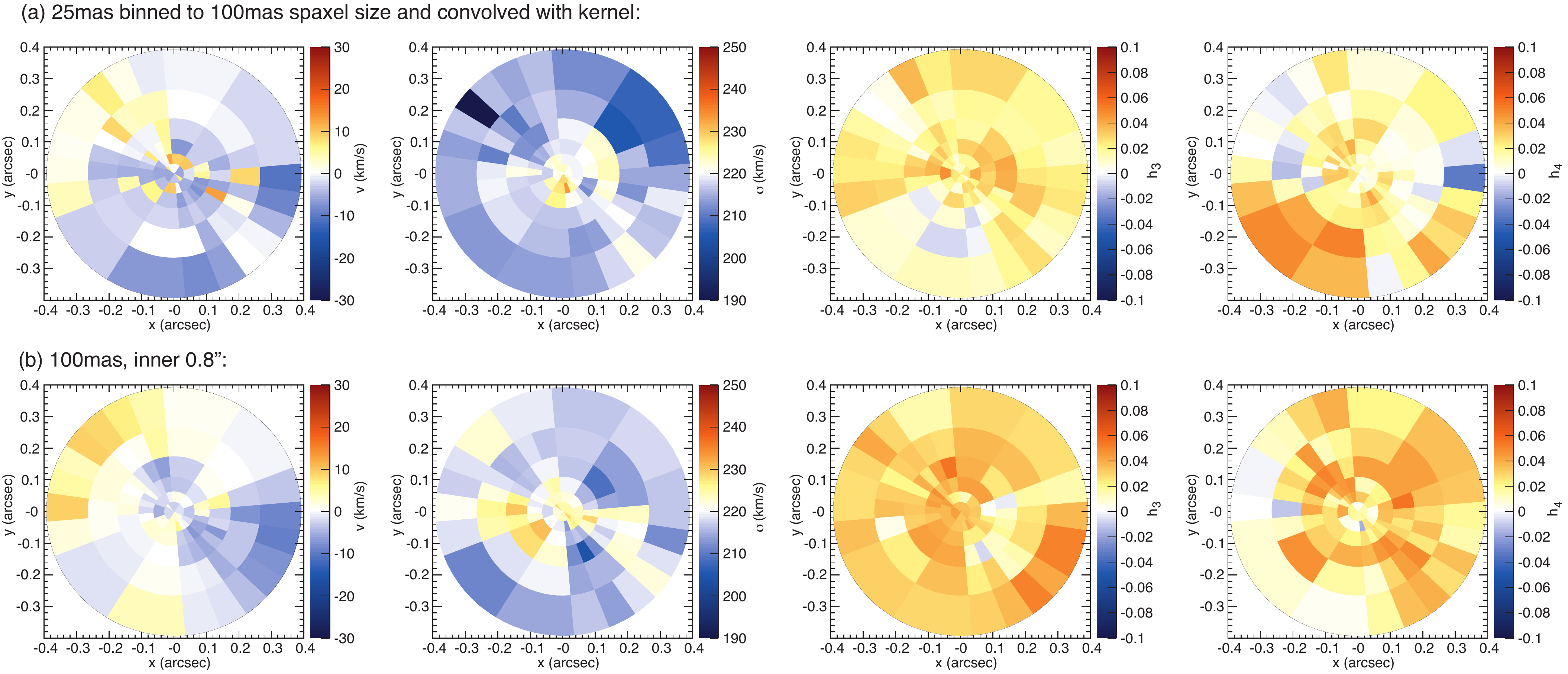}
  \caption{The 25mas data convolved with the kernel from
  Fig. \ref{fig:2}c (upper row, (a)), compared to the central part of
  the 100mas kinematics (lower row, (b)).}\label{fig:7}
\end{minipage}
\end{figure*}

In addition to the difficulty in explaining the structures of $v$ and
$\sigma$ in the centre, it is also surprising that neither the $v$ nor
the $\sigma$ peak can be seen in the 100mas data. We investigated if
this is due to resolution in the following way: The 25mas data were
binned $4\times4$ to get the same spaxel sizes as in the 100mas
scale. Then the binned data were convolved with the kernel
(Fig. \ref{fig:2}c) that transforms the 25mas SINFONI image of
Fornax\,A into the 100mas image, before being binned according to the
radial and angular scheme described above. Fig. \ref{fig:7}a
shows the resulting 2D kinematic maps compared to the inner
$0.8$~arcsec of the 100mas data (Fig. \ref{fig:7}b). The
structures seen in the 25mas data mostly disappeared after the
convolution and the convolved 25mas kinematics now look very similar
to the 100mas kinematics.

The interpretation of the features seen in the unconvolved 25mas data
is more difficult. In the following possible explanations are
discussed.

\subsubsection{Technical aspects}
The features detected in the 25mas data, especially the central
velocity rise, are not due to template mismatching, as they remain
also when other template combinations or single templates are used. Of
the nine 25mas templates, two (a K1{\sc V} and a K3{\sc V} star with CO EWs
$\la5\mathrm{\AA}$) cannot fit the data at all. With each of the other
templates the kinematics of Fig. \ref{fig:6}a can be
reproduced at least qualitatively independently of the specific choice
of the smoothing parameter.

The results can also be reproduced when fitting slightly different
wavelength regions, as long as the CO bandheads are
included. Measuring the kinematics from the other absorption lines
(\cai, Fe\,{\sc I} and \mgi) does not work because of the weakness of the lines
compared to CO. Here a much higher S/N would be required.

The central bins are very small and only cover a limited number of 
spaxels. Therefore one or two spaxels alone could cause the increase of
$v$, because \textrm{e.g.} of an incorrect sky subtraction or bad pixel
correction in the CO bandhead region of these spaxels. In addition, as
the S/N of the single spaxels decreases relatively slowly from the
centre to the outer parts, the S/N of the central bins is lower
($\sim50$) than at larger radii ($60-80$), so the errors are larger
and the large velocity could be just a statistical fluctuation which
would occur at larger radii as well if the S/N was equally low. A
single-spaxel effect can be in principle verified or excluded by
simply fitting the unbinned spectra. The problem is, that the S/N is
around or below $30$ which does not allow a reliable measurement of
the kinematics. We first used MPL with a large $\alpha$ and then, as a
consistency check, fitted simple Gaussians to the unbinned
spectra. A central $v$ peak is present in both cases.

\subsubsection{A central disc or star cluster}

Central velocity dispersion drops are common in spiral galaxies
(\textrm{e.g.} \citealt{Peletier-07}) and usually associated with central
discs. Central discs can be formed from gas inflow towards the centre
and subsequent star formation \citep{Wozniak-03}. In early-type
galaxies $\sigma$-drops are rarely found. NGC\,1399 is the only case
where the observed central $\sigma$-drop is discussed in detail
\citep{Houghton-06,Gebhardt-07,Lyubenova-08}. \citet{Lyubenova-08}
conclude that it is a dynamically distinct subsystem with a different
stellar population.  

Fornax\,A is not a clearly elliptical galaxy, but can be classified as
a ``peculiar'' galaxy, as some remaining features like shells and
ripples from a major gas-rich merger about $3$~Gyr ago are still
clearly visible \citep{Schweizer-80,Schweizer-81,Goudfrooij-01}. There
are also hints that a minor merger with another gas-rich galaxy
happened $\sim0.5$~Gyr ago \citep{Mackie-98}. Thus it would not be
surprising if remainders of the merging processes were also found in
the centre, like a disc produced from merger-triggered gas inflow or
an infalling disc- or cluster-like central part of one of the merging
components. A nuclear disc could explain the relatively thin
($\la8$~pc) and apparently elongated structure of the
$\sigma$-drop. With a width of only $\la8$~pc this is probably the
thinnest $\sigma$-drop ever detected. It is even narrower than the
drop in NGC\,1399 and also far less extended than the typical drops in
spiral galaxies. As its size is comparable to the spatial resolution
it is possibly unresolved. On the other hand rotation would be seen
if there really was a disc.

The centre could still be unrelaxed and in the process of
merging. This could induce signatures in the kinematics like those
seen in Fornax\,A. Simulations of unequal-mass mergers with a SMBH in
the primary galaxy and a significantly smaller secondary galaxy
(without black hole) were performed by \citet{Holley-00}. Based on
this it depends critically on the orbital decay trajectory of the
secondary, and also somewhat on the black-hole mass and on the mass
ratio of the galaxies, whether the secondary galaxy is destroyed by
the merger or not.  The time-scale for a stellar cluster falling in on
a purely radial orbit -- also suggested by \citet{Gebhardt-07} as one
possible explanation for the $\sigma$-drop in NGC\,1399 -- however
seems to be much shorter (of the order of about $10^7$~yr) than the
date of the last merger ($\sim10^9$~yr), if the conditions in
Fornax\,A are similar as in \citet{Holley-00}.  For nonradial mergers
the formation of a rotating stellar disc is possible. As we do not
know any of the details of the merger history with desirable precision
for Fornax\,A, and as the time-scales seem to vary depending on the
initial parameters of the merger simulation, it is impossible to tell
whether the centre is still unrelaxed or not.

\subsubsection{The effect of dust}
Fornax\,A contains a lot of dust features (see lower panel of
Fig. \ref{fig:3}) and this dust might alter the
kinematics. On the other hand observations in the $K$ band should
minimize the effects of dust and indeed already the uncorrected SOFI
\emph{Ks} band image shows only very little amounts of dust compared
to optical \emph{HST} images. Some dust is located close to the
centre, but not in the central $\sim2$~arcsec. \citet{Shaya-96} find a
low colour excess in the central pixels of an \emph{HST} WFPC $V-I$
image and conclude that here the dust extinction is only very
light. Consistently the SINFONI 25mas and 100mas images also do not
show any signs of dust (\textrm{cf.} Fig. \ref{fig:1}). The 250mas
image shows hints of dust in the SE region, spatially coincident with
the high-$\sigma$ region just outside the 100mas field of
view. Therefore the dust outside the central $2-3$~arcsec could be the
reason for the larger average $\sigma$ of the 250mas data, in
agreement with the findings of \citet{Silge-03}.

Due to the lack of dust in the 25mas image dust does not seem to be a
plausible explanation for the central $\sigma$ structure. However the
scale of the effect of dust on the kinematics might depend on the
location of the dust along the line of sight. \textrm{E.g.} a certain
amount of dust directly behind the centre could have a noticeable
effect without being noticed in the photometry at the given S/N.  To
further substantiate this we tested the effects on the kinematics when
dust obscures light from stars behind or in front of the centre of a
galaxy using an N-body model realisation of the best-fitting dynamical
model (\textrm{cf.} \S 5.2.3) as described in \citet{Thomas-07}. We
calculated the kinematics of the model by obscuring the central part
at different radii and with different opacities and found that only a
noticeable decrease in $\sigma$ can be induced by putting a very thick
layer of dust in front of the centre, but in this case the surface
brightness would be reduced by more than a magnitude. Therefore we can
exclude that the structure of the 25mas kinematic fields is produced
by dust.

\subsubsection{Stellar populations}
\begin{figure*}
\begin{minipage}{168mm}
\centering
\includegraphics[width=\linewidth,keepaspectratio]{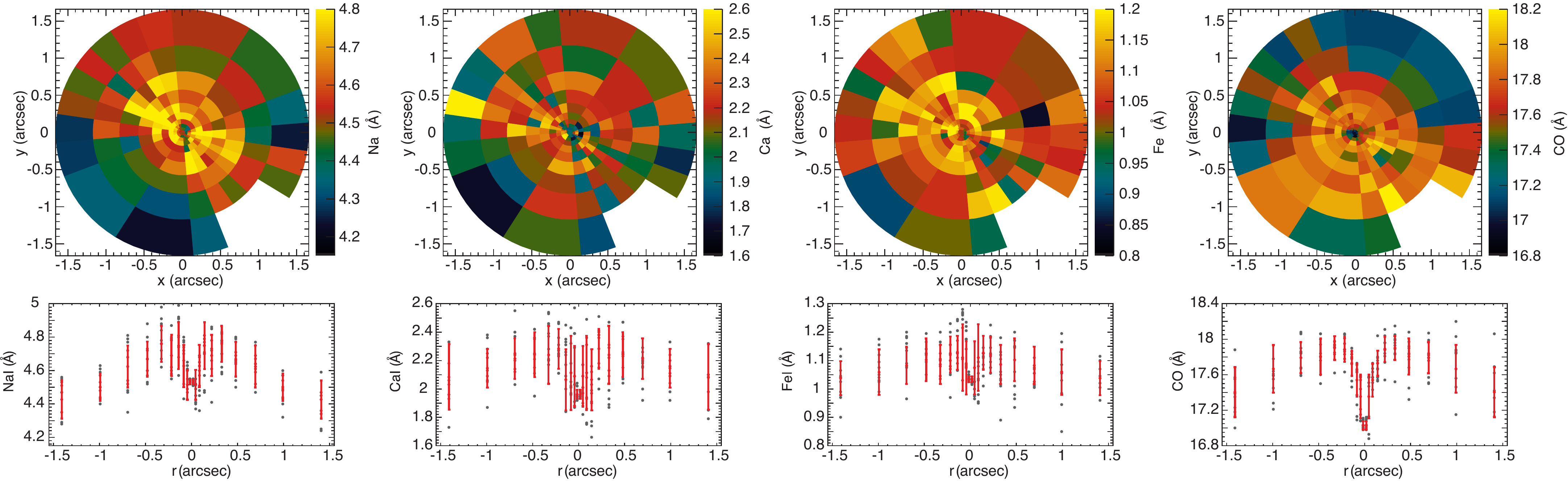}
  \caption{\nai, \cai, \fei\ and CO indices of Fornax\,A in the
  central $1.5$~arcsec (100mas scale). The upper row shows the 2D
  index fields, below the position-$i$ diagrams are shown (where $i$
  is one of the four indices). The individual values are plotted in
  grey, while overplotted in red are the mean indices and their rms of
  the bins belonging to a ring of radius $r$.}\label{fig:8}
\end{minipage}
\end{figure*}

Another explanation might be that the stellar population is different
in the central region. Blue stars in the nucleus could in principle
also be responsible for the low colour excess \citet{Shaya-96}
find. The $K$ band absorption line equivalent widths for Fornax\,A are
presented in \S 4 in detail. The 100mas maps of the absorption line
indices (Fig. \ref{fig:8}) clearly show that the line indices
change within the central $\sim0.4$~arcsec. All indices show a
decrease in this region, CO as the deepest absorption feature shows it
most clearly. Unfortunately this trend is not as clear in the 25mas
indices, as here the S/N is only about half as in the 100mas data and
the scatter therefore is much larger.

Thus a dynamically colder subsystem with accordingly different line
indices could be an explanation of the distorted kinematics, but the
S/N of the 25mas data unfortunately is not large enough to resolve any
spatial coincidence with the $v$ peak or $\sigma$ drop. Similar
conclusions have been made by \citet{Lyubenova-08} for NGC\,1399,
where the central $\sigma$-drop coincides with a drop of the \nai\ and
CO indices.

If a changing stellar population in the central region has an effect
on the kinematics or not was tested with the N-body model mentioned in
the previous section. Making the stars in the central $0.1$~arcsec ten
times brighter (corresponding to a ten times lower mass-to-light
ratio) results only in very small changes of $\Delta\sigma=4$~\kms\
and $\Delta h_{4}=-0.002$, both within the errors of the data and much
smaller than the observed variations. $100$ times brighter stars would
produce noticeable changes of $\Delta\sigma=25$~\kms\ and $\Delta
h_{4}=-0.015$. Making the stars darker does not have an effect at all
on the kinematics. As there is no photometric evidence of a very
bright star cluster in the centre, we conclude that a stellar
population alone, which is different from the surrounding population,
is not the reason for the disturbed central kinematics.

\subsubsection{AGN emission}
A decrease of the line indices would also be expected for AGN, where
the absorption features are diluted by the non-stellar emission of the
active nucleus. In this case also the measurement of $v$ and $\sigma$
could be affected. As shown in \citet{Davies-07b} the PSF can be
reconstructed from the decrease of the CO equivalent width at the
position of the point source.  Comparing the PSF measured this way
with the measured or the noisy reconstructed PSFs from
Fig. \ref{fig:2} is difficult, because (1) we would need to measure
the CO index of the unbinned spectra, which have a lower S/N than the
binned ones and thus errors larger than the small
$\sim1.0~\mathrm{\AA}$ decrease observed in the binned spectra, and
(2) for the higher S/N 100mas exposure we do not exactly know the true
PSF. From the binned spectra we cannot deduce the two-dimensional
PSF. A rough estimate of the FWHM of the CO dip can be obtained from
fitting the averaged CO indices of a ring (red points in
Fig. \ref{fig:8}). The resulting FWHM of $0.17$~arcsec in the
100mas scale is, considering the large errors introduced by the
binning, well in agreement with the measured PSF. This would favour
the AGN scenario.

\begin{figure}
\centering
\includegraphics[width=.9\linewidth,keepaspectratio]{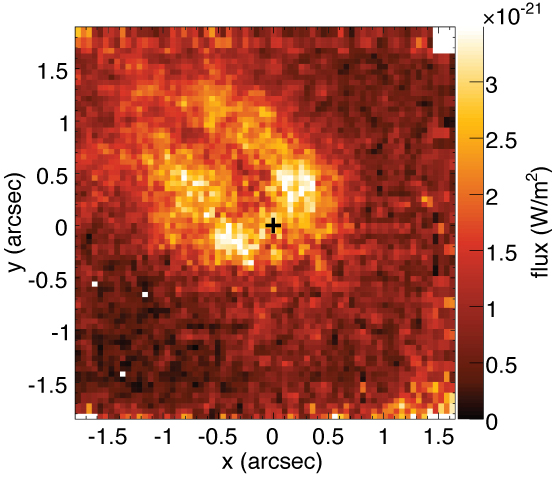}
  \caption{H$_{2}$ emission of Fornax\,A (100mas scale). The photometric centre is marked by a cross.}\label{fig:9}
\end{figure}

It was found by \citet{Iyomoto-98} that the AGN in Fornax\,A, which
powered the very luminous radio lobes, became dormant during the last
$0.1$~Gyr. Still the nucleus shows weak radio \citep{Geldzahler-84}
and weak X-ray emission with a spectrum typical for a low-luminosity
AGN \citep{Kim-03} and is bright in the UV \citep{Fabbiano-94}. In
high-S/N optical spectra several emission lines are present ([OIII],
H$\beta$, NI), but very weak \citep{Beuing-02}. In the SINFONI spectra
there are no apparent emission lines typical for an AGN present at the
given S/N. Only weak molecular hydrogen ($1-0$S(1) H$_2$ at
$\lambda=2.122~\umu$m) with a S/N around $5$ is found in a region
north-east of the nucleus (see Fig. \ref{fig:9}), but does not
seem to be associated with the central source. The total H$_2$ flux in
the 100mas field of view is $5.65\times10^{-18}$~W~m$^{-2}$ and in a
$3$~arcsec aperture centred on the continuum peak it is
$3.64\times10^{-18}$~W~m$^{-2}$. The H$_2$ could be excited by X-rays,
as a similar northeast-southwest elongation is present in Chandra
data. This X-ray emission was associated with hot inter-stellar medium
with a temperature of $kT=0.62$~keV by \citet{Kim-03}.

A weak AGN present in Fornax\,A would also become noticeable by a
slight change of the continuum slope, making the nucleus redder. This
is due to the AGN UV-continuum which heats surrounding dust that emits
in the near-IR \citep{Oliva-99}. We indeed find that the slope of the
continuum changes slightly in the inner $\approx0.1$~arcsec by no
more than $2-3$~per~cent. 

\begin{figure*}
\begin{minipage}{168mm}
\centering
\includegraphics[width=\linewidth,keepaspectratio]{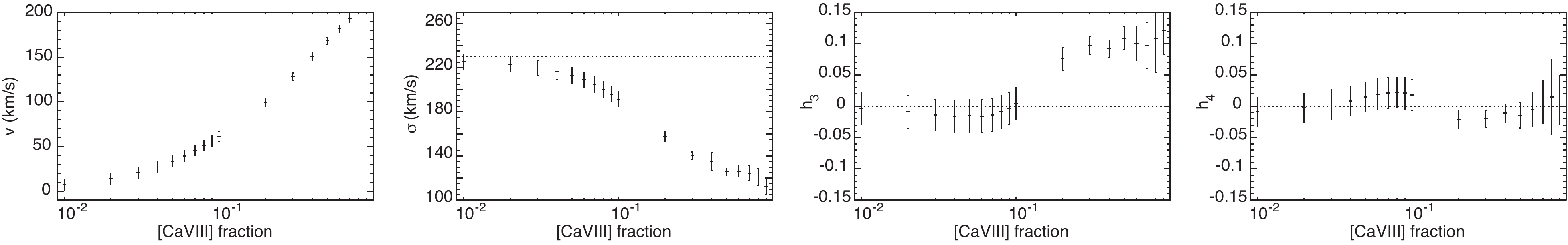}
  \caption{Mean $v$, $\sigma$, $h_3$ and $h_4$ as a function of \caviii\ contribution. The error bars are the $67$~per~cent intervals obtained from fitting $100$ realisations of the broadened template spectrum with added noise such that the S/N is similar to the central Fornax\,A spectra.}\label{fig:10}
\end{minipage}
\end{figure*}

\begin{figure*}
\begin{minipage}{168mm}
\centering
\includegraphics[width=\linewidth,keepaspectratio]{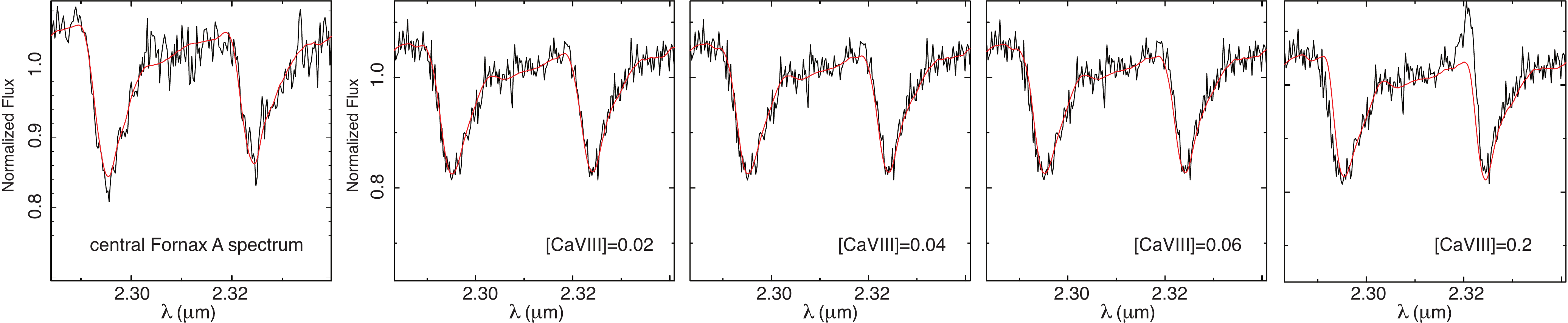}
  \caption{Fit (red) to broadened template star spectra (black) with different \caviii\ contributions compared to the fit to the central 25mas Fornax\,A spectrum with the largest measured velocity (leftmost plot).}\label{fig:11}
\end{minipage}
\end{figure*}

\begin{figure}
\centering
\includegraphics[width=\linewidth,keepaspectratio]{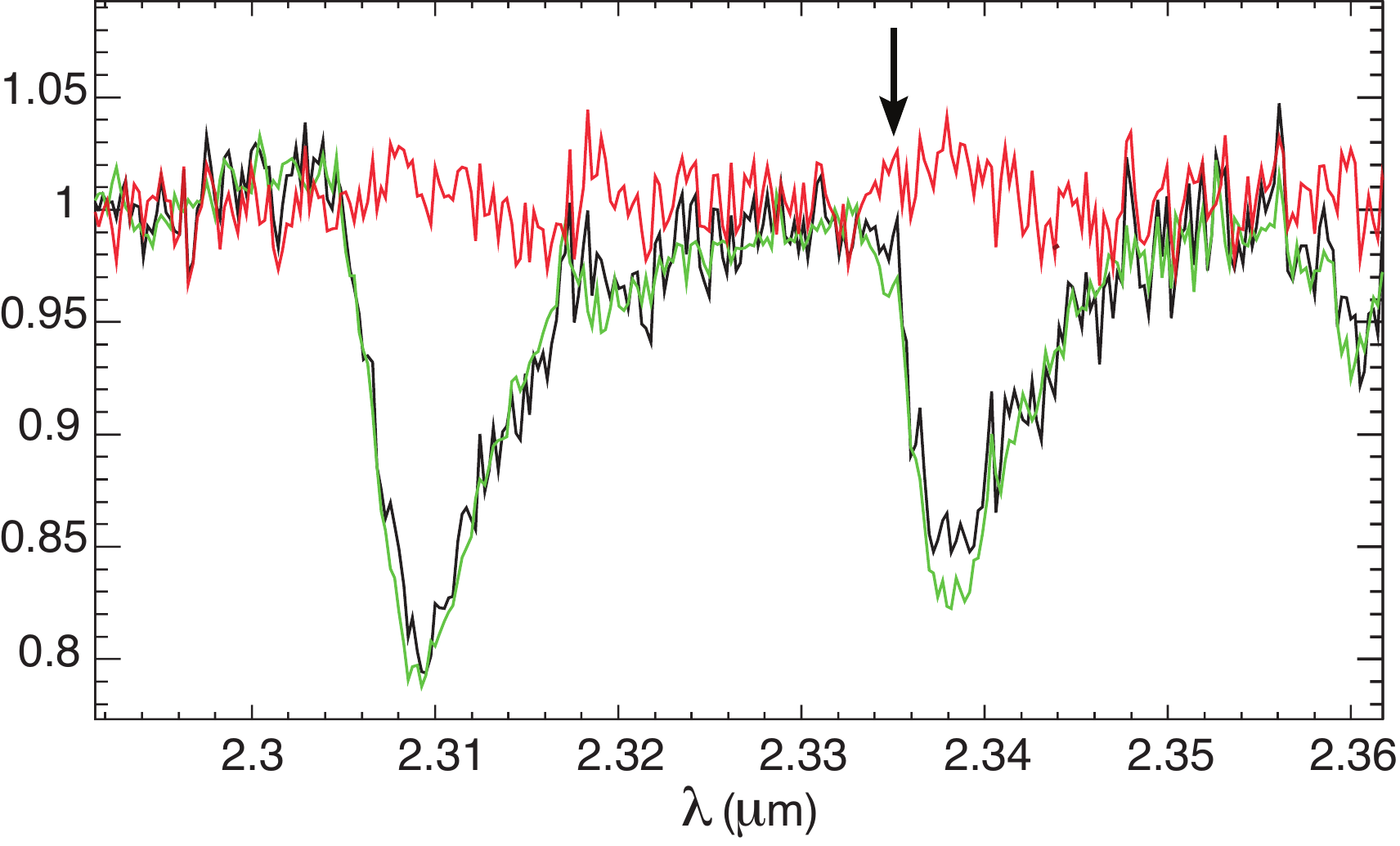}
  \caption{Combined spectra of the central region (black) at radii $r<0.03$~arcsec and of a ring at larger radii $0.11<r<0.17$~arcsec (green). In red the residuals between the two spectra are plotted, shifted by $+1$. The position where the \caviii\ line is assumed to be is marked with an arrow.}\label{fig:12}
\end{figure}

Some AGN show a certain coronal emission line in the region of the CO
bandheads, which alters the measured kinematics: \caviii\ at
$2.3213~\umu$m \citep{Portilla-08,Davies-06}. As it is located
directly at the left edge of the second CO bandhead, already a very
small contribution can have a significant effect on the measured
kinematics.  We tested this using two different approaches. First we
tried to measure the kinematics using only the first bandhead. The
velocity map in this case has a much larger scatter and shows many
peaks of $20$-$30$~\kms\ in the field of view. $\sigma$ remains
largely unchanged. When the first together with the third and fourth
bandhead are used, the velocity map is smooth and shows no increase in
the centre. The $\sigma$ map is qualitatively similar to the one
measured with the first two bandheads, but the absolute value of
$\sigma$ is much larger. The third and fourth bandheads are
significantly distorted by sky emission lines and therefore do not
give reliable results. The second approach was to use a standard star,
convolve it with a LOSVD ($v=0$~\kms, $\sigma=230$~\kms, $h_3=h_4=0$)
and add the \caviii\ line with a typical FWHM of $150$~\kms.  The
results are shown in Figs. \ref{fig:10} and \ref{fig:11}.  A
\caviii\ contribution of $\la4$~per~cent would not be seen in the
spectra or the fit at the given S/N, but can increase $v$ by
$\la25-30$~\kms\ and decrease $\sigma$ by $\la10$~\kms\ ($\sigma$
measured with zero \caviii\ contribution is already a few \kms\ lower
than the real $\sigma$, see \textrm{e.g.} Fig. \ref{fig:B1} in
Appendix B). Therefore a weak AGN with some \caviii\ contribution is
the most logical explanation for the observed velocity increase. The
observed $\sigma$ decrease probably has for the most part a different
origin (most probably a stellar population effect), as it is
significantly larger than the $\la10$~\kms\ estimated from the
simulations. In addition the high-$v$ region only partly coincides
with the low-$\sigma$ region. The hint of a \caviii\ line signature appears when stacking all spectra
within the region covered by the innermost bins and comparing it to a
combination of all spectra within a ring further out
($0.11<r<0.17$~arcsec; see Fig. \ref{fig:12}).

As accurate kinematics for the centre cannot be obtained from only one
bandhead or by including the third and/or fourth bandhead, we will
exclude the central bins in some of the the dynamical models. As the
black hole sphere of influence is large, this solution will not
degrade the reliability of the resulting black-hole mass
significantly.


\section{Line indices}

The optical line strengths of Fornax\,A have been analysed extensively
by \citet{Kuntschner-phd,Kuntschner-00}. They found a young (around
$2$~Gyr) and very metal-rich stellar population. The optical index
gradients slightly decrease with radius. Unfortunately their data have a
resolution of only $1$~arcsec, therefore they do not resolve the
structure in the central $0.4$~arcsec. The age of a number of globular
clusters in Fornax\,A was found to be about $3$~Gyr
\citep{Goudfrooij-01b}. Thus a plausible scenario is that the globular
clusters formed during the last major merger event $3$~Gyr ago
\citep{Goudfrooij-01b} and that the young stellar population formed
from infalling molecular gas also as a result of this merger
\citep{Horellou-01}.

The spectral features in the $K$ band can in principle be likewise
interpreted in terms of age and metallicity. Unfortunately there is
not such a sophisticated theoretical spectral synthesis model
available yet as for optical indices and studies of $K$ band indices
are rare and often include only small samples. A recent study of
\citet{Silva-08} investigates the behaviour of the $K$ band indices of
cluster stars of known age and metallicity and a sample of Fornax
galaxies. Their published line indices of Fornax\,A are derived by
extracting the mean spectrum over an aperture of
$2\times3$~arcsec from our SINFONI 100mas data.

In Fig. \ref{fig:8} we show the two-dimensional $K$ band line index
maps (\nai, \cai, \fei\ and CO) of Fornax\,A for the 100mas scale,
because the S/N is largest for this scale. The index definitions were
taken from table 6 in \citet{Silva-08}. Instead of \feia\ and \feib\
we use \fei=(\feia + \feib)/2, as they are highly correlated. The
definition of the CO index is different from the definition of the CO
equivalent width used above \citep{Silge-03}. Here the continuum is
defined by four blue continuum bands instead of a blue and a red
one. The continuum on the red side is diminished by the CO absorption
bands, therefore an accurate measurement of the CO index is possible
by using blue continua only. Using the definition of \citet{Silge-03}
nevertheless is sufficient to decide quickly which templates to use
and to notice any particularities (see Table \ref{tab:templates}
for a comparison between the CO EWs of the kinematic template stars
derived by the two different index definitions). As in
\citet{Silva-08} we broadened the SINFONI spectra to match the ISAAC
resolution of their data and to be able to directly compare our
results to their findings.

\begin{table}
 \centering
  \caption{Mean near-IR line indices of Fornax\,A.}\label{tab:indices}
  \begin{tabular}{lcc}
  \hline
   &  this work & \citet{Silva-08}\\
 \hline
 \nai\   & $4.61$ ($0.15$) &  $4.70$ ($0.14$)  \\
 \cai\   & $2.14$ ($0.21$) & $2.48$ ($0.14$)   \\
 \feia\   & $1.44$ ($0.11$) & $1.491$ ($0.066$)   \\
 \feib\   & $0.73$ ($0.09$) & $0.843$ ($0.065$)   \\
 CO      & $17.61$ ($0.33$) & $17.53$ ($0.30$)  \\
\hline
\end{tabular}

\medskip
Note. The indices are given in units of $\mathrm{\AA}$. The middle
column gives the near-IR line indices determined by averaging the
values shown in Fig. \ref{fig:8} (upper row) and the according rms
errors. The indices determined by \citet{Silva-08} are given in the
right column with their errors.
\end{table}

%binning anders, leicht anderes fov, ca und fe much weaker than CO and Na
The measured indices, averaged over the entire field of view, are in
good agreement with the measurements of \citet{Silva-08}, as shown
in Tab. \ref{tab:indices}. In two dimensions, all indices show the same behaviour: a more
or less clear dip in the centre and further out a negative
gradient. This is most obvious in CO and \nai.  Negative gradients are
also present in the optical indices and it is therefore reasonable to
assume that the $K$ band index gradients can be likewise interpreted
as age or metallicity gradients. \citet{Kuntschner-phd} find for the
Fornax ellipticals that the index gradients are linked to a
metallicity gradient, and not an age gradient. They assume that the
negative gradients in Fornax\,A are likewise due to a metallicity
gradient only, but their data do not go very far out in radius and
therefore they cannot draw secure conclusions. \citet{Marmol-08}
quantify a metallicity dependence of the CO index, which supports this
interpretation.

What causes the dip in the very centre is less obvious. It might be
one of the scenarios discussed above -- AGN emission, dust or a
different stellar population like a nuclear cluster or a population of
younger, bluer stars of a post-starburst
population. \citet{Fabbiano-94} detected a UV-bright unresolved
central source with $r<3$~pc, which according to them could be
explained \textrm{e.g.} by a cluster of O stars or an AGN. The FWHM of
the CO dip is very similar to the spatial resolution and thus
unresolved, corresponding to an upper limit in size of
$r\la4$~pc. Thus it is likely that the UV emission and the CO dip are
caused by the same structure. Both early-type stars -- due to the lack
of typical late-type star absorption lines -- and non-thermal emission
would cause the observed decrease of all indices. Br\,$\gamma$
absorption is the only strong feature in early-type $K$ band stellar
spectra, Br\,$\gamma$ emission is observed in many galaxies with
nuclear activity or star formation, thus this feature could be used to
distinguish between these two possibilities. In the central SINFONI
spectra we do not detect Br\,$\gamma$ neither in emission nor in
absorption. Thus emission or absorption is either not present, too
weak to be detected, or Br\,$\gamma$ emission and absorption just
compensate. \caviii\ emission caused by the AGN is also very well
hidden (see Fig. \ref{fig:12} and discussion in \S 3.4.5). Due
to the broader PSF its influence is eliminated in the 100mas
data. Other emission lines typically present in AGN spectra are absent
as well in the central region. Weak H$_{2}$ emission is present in the
north-eastern region, but not correlated to the central source (see
Fig. \ref{fig:9}). In order to resolve the reasons that cause
the central line index dip and kinematics, very high S/N observations
with a better spatial resolution as well as self-consistent spectral
synthesis models for the interpretation of the $K$ band indices are
needed.

\section{Dynamical models}

Based on the $M_\bullet$-$\sigma$ relation of \citet{Tremaine-02} and
our $\sigma$ measurements a black hole with a mass around
$2\times10^8$~M$_\odot$ would be expected. Earlier models by
\citet{Shaya-96} suggest a ten times higher black-hole mass, but they
rely on larger $\sigma$ values. \citet{Davies-00} observed Fornax\,A
with the MPE 3D IFS \citep{Weitzel-96} and measured $v$ and $\sigma$
using the CO bandheads at $2.2$~$\umu$m. Their $\sigma$ is lower than
the values used by \citet{Shaya-96} and comparable to our
values. Together with the models of \citet{Shaya-96} they obtain a
SMBH mass of $\la10^{9}$~M$_{\odot}$. However, they cannot exclude that
radial anisotropy may account for this large mass. A lower limit of
the central mass concentration of $75$~M$_{\odot}$ was estimated by
\citet{Fabbiano-94} based on the measured nuclear UV emission.

We used the axisymmetric code of \citet{Gebhardt-00c,Gebhardt-03} in
the version of \citet{Thomas-04} to determine the mass of the SMBH in
Fornax\,A. It is based on the \citet{Schwarzschild-79} orbit
superposition technique and comprises following steps: (1) Calculation
of the gravitational potential of the galaxy from the stellar mass
density $\rho_*$ using a trial black-hole mass $M_{\bullet}$ and
mass-to-light ratio $\Upsilon$. (2) Generation of an orbit library for
this potential and construction of a weighted superposition of orbits
that best matches the observational constraints. (3) Repetition of the
first two steps with different values for $\Upsilon$ and $M_{\bullet}$
until the eligible parameter space is systematically sampled. The
best-fitting parameters then follow from a $\chi^2$-analysis. Our
orbit libraries contain around $2\times7000$ orbits. The deprojected
luminosity density is a boundary condition and hence exactly
reproduced, while the LOSVDs are fitted in $25$ velocity bins between
$-880\dots+880$~\kms\ with a bin width of $\approx75$~\kms. Special
care was taken when implementing the PSF. Due to its asymmetric shape
the PSF was not fitted, but the models were rather convolved with the
two-dimensional image of the PSF reference star directly (\textrm{cf.}
\citealt{Nowak-07}). All modelling was done with minimal
regularisation.

A big advantage of integral-field data compared to longslit data only
is the assurance of whether the assumption of axisymmetry is
legitimate by comparing both the kinematics and the results of the
dynamical modelling of all four quadrants. In case all quadrants
produce the same results, the data can be folded and averaged, so that
the errors are reduced. This results in a very large number of
models. For Fornax\,A we calculated around $5000$ models in total,
covering a broad parameter space in $M_{\bullet}$
($0\dots5\times10^8$~M$_\odot$) and $\Upsilon$ ($0.55\dots0.95$) for
each quadrant, several combinations of datasets and PSFs using an
inclination of $90\degr$. At the beginning we computed some test
models using either the longslit data only (\S 5.1.1) or the SINFONI
data only (\S 5.1.2) in order to constrain the $\Upsilon$ range.  As a
second step each of the four quadrants was modelled separately using
the longslit data and either the 25mas data (\S 5.2.1) or the 100mas
data (\S 5.2.2) or a combination of the two (\S 5.2.3). This step is
essential as it reveals possible inconsistencies between the datasets,
non-axisymmetries and other problems. The assumed diameter of the
sphere of influence is around $0.46$~arcsec, resolved by both SINFONI
datasets. The 250mas data were not included in the modelling, as they
do neither resolve the sphere of influence nor do they add
significantly more information in the outer parts that would justify
the enhanced amount of computing time. After a careful examination of
the results and verifying the consistency of the data with
axisymmetry, the four quadrants were averaged and modelled for the
same combination of datasets as for the single quadrants.

\subsection{The stellar dynamical \emph{Ks} band mass-to-light ratio $\Upsilon$}
In order to check if the stellar mass-to-light ratios $\Upsilon$
obtained for individual datasets, agree and if they
are consistent with $\Upsilon$ from population synthesis models
\citep{Maraston-98,Maraston-05}, we determine $\Upsilon$ by modelling
single datasets.

\subsubsection{$\Upsilon$ from longslit data}

\begin{figure}
\centering
\includegraphics[width=\linewidth,keepaspectratio]{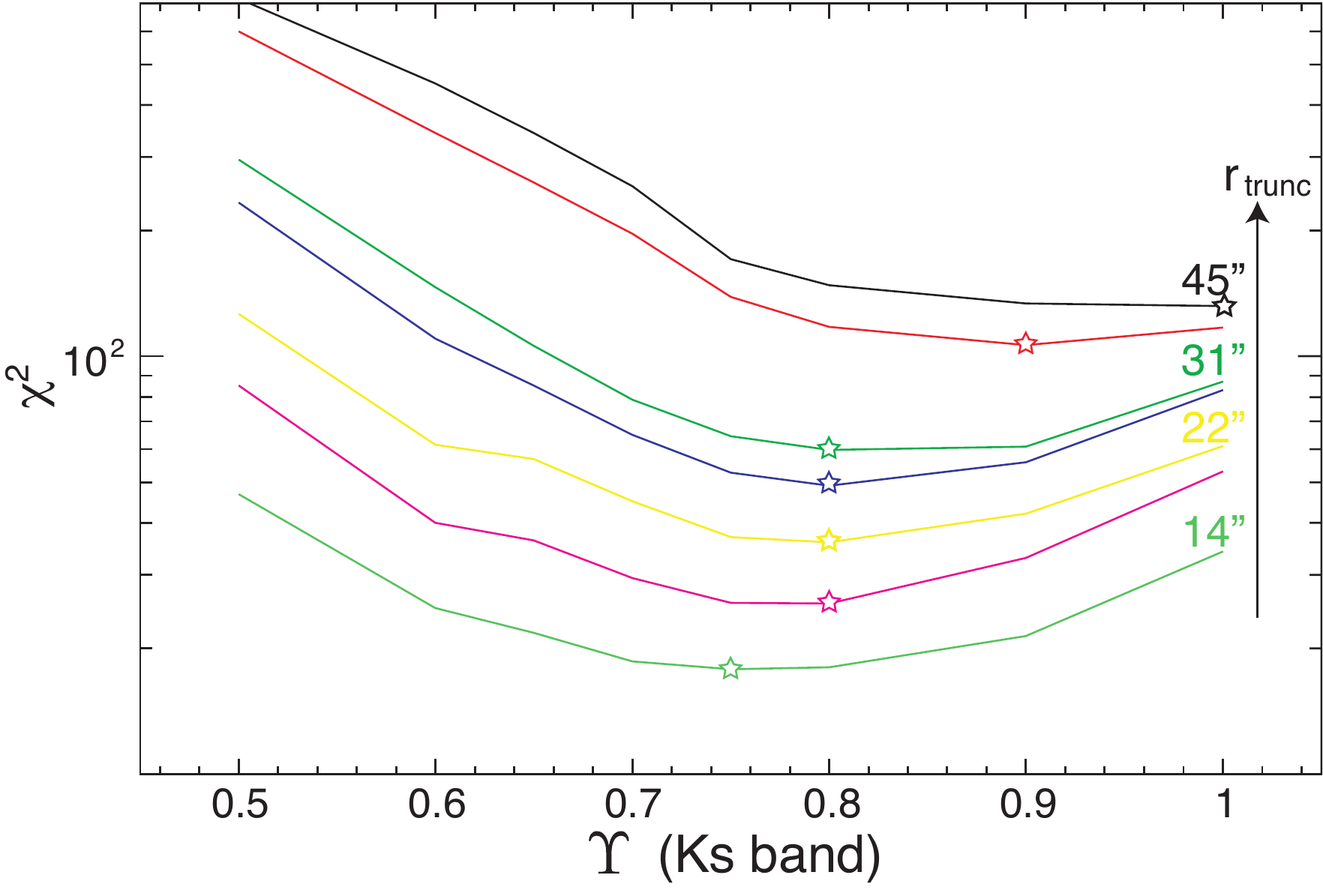}
\caption{$\chi^2$ as a function of mass-to-light ratio $\Upsilon$
for models calculated with the longslit data alone, truncated at
different radii $r_\mathrm{trunc}$ and without black hole. The models
with the smallest $\chi^2$ are marked with a star.}\label{fig:13}
\end{figure} 

First a number of models were calculated using exclusively the
longslit data without black hole covering a broad range in $\Upsilon$
(in units of
$\mathrm{M}_{\odot}/\mathrm{L}_{\odot}^{\mathrm{\emph{Ks}}}$). The
longslit data were truncated at different radii $r_{\mathrm{trunc}}$
between $14$~arcsec and $45$~arcsec. The resulting best-fitting
$\Upsilon$ (Fig. \ref{fig:13}) is roughly constant with
$r_{\mathrm{trunc}}$ and $\approx0.75-0.8$ for longslit data at
$r_{\mathrm{trunc}}\la31$~arcsec$\approx2.8$~kpc, which roughly
corresponds to the effective radius of Fornax\,A
($R_{\mathrm{e}}=36^{+5.7}_{-11.2}$~arcsec, \citealt{Bedregal-06}). At
larger truncation radii $\Upsilon$ increases strongly, a sign that
here the dark halo starts to have an effect. This is consistent with
the results of \citet{Thomas-07b}, who found dark matter fractions of
$\approx30$~per~cent at $3$~kpc in similar bright Coma galaxies.  Thus
in the following only longslit data at $r<31$~arcsec will be used for
the dynamical modelling.

\subsubsection{$\Upsilon$ from SINFONI data}

\begin{figure}
\centering
\includegraphics[width=\linewidth,keepaspectratio]{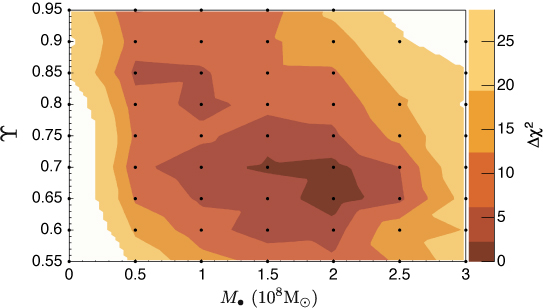}
\caption{$\Delta\chi_0^2=\chi^2-\chi_\mathrm{min}^2$ as a function of
$\Upsilon$ (\emph{Ks} band) for models calculated with the SINFONI 100mas data alone
and a small number of black-hole masses $M_\bullet$. The coloured
regions are the $1\dots5\sigma$ errors in $\Delta\chi_{0}^{2}$. Each black
dot represents a model.}\label{fig:14}
\end{figure} 

For the SINFONI 100mas data alone the best-fitting $\Upsilon$ is
slightly lower ($\approx0.7$ in units of
$\mathrm{M}_{\odot}/\mathrm{L}_{\odot}^{\mathrm{\emph{Ks}}}$, see
Fig. \ref{fig:14}), but within the errors consistent with
$\Upsilon$ derived from the longslit data alone. The contours are
strongly elongated towards large $\Upsilon$ with slightly decreasing
$M_\bullet$, as large $\Upsilon$ in the centre can somewhat compensate
a lower $M_\bullet$. This degeneracy between $M_\bullet$ and
$\Upsilon$ is even stronger for the tiny field of view 25mas data
alone, where $\Upsilon$ cannot be constrained very well. This
emphasizes the importance of including the longslit data in the
dynamical modelling, which makes the degeneracy virtually disappear.

In order to be able to compare our stellar dynamical \emph{Ks} band
$\Upsilon$ to predictions from population synthesis models, we need to
scale it to units of $\mathrm{M}_{\odot}/\mathrm{L}_{\odot}^{K}$ by
multiplying it with the ratio of the bandwidths:
\[
\Upsilon_K=\Upsilon\frac{\Delta\mathrm{\emph{Ks}}}{\Delta
K}=\Upsilon\frac{0.275~\umu\mathrm{m}}{0.390~\umu\mathrm{m}}\approx0.5
\]
For a $2-3$~Gyr old population with a high metallicity
$[Z/H]=0.3-0.45$ \citep{Kuntschner-phd,Kuntschner-00} the models of
\citet{Maraston-98,Maraston-05} predict $\Upsilon_K\approx0.3\dots0.6$
(Salpeter IMF) and $\Upsilon_K\approx0.2\dots0.4$ (Kroupa IMF). Thus
the measured stellar dynamical $\Upsilon$ is in good agreement with a
Salpeter IMF but larger than predicted for a Kroupa IMF. When taking
into account the error in the measured dynamical $\Upsilon$ and a
possible error in the distance assumed for Fornax\,A, then the
dynamical $\Upsilon$ is marginally in agreement also with the Kroupa
IMF. \citet{Cappellari-06} also find that the dynamical $\Upsilon$ for
E/S0 galaxies are in agreement or larger than the $\Upsilon$ predicted
using a Kroupa IMF.

\subsection{The black-hole mass $M_{\bullet}$}

\subsubsection{$M_{\bullet}$ from the 25mas and the longslit data}

After having constrained the possible $\Upsilon$ range we now focus on
the determination of the mass of the central black hole. We modelled
the 25mas dataset together with the longslit data for each quadrant
separately and at an inclination of $90\degr$. The resulting values
for $M_{\bullet}$ and $\Upsilon$ with the $3\sigma$ error bars are
listed in Tab. \ref{tab:results}. The best-fitting black-hole masses
and mass-to-light ratios are in the range
$M_{\bullet}=(2.0\dots3.25)\times10^{8}$~M$_{\odot}$ and
$\Upsilon=0.675\dots0.775$ and agree very well within $2\sigma$
($\Delta\chi_0^2=\chi^2-\chi^2_{\mathrm{min}}=6.2$) errors.

\subsubsection{$M_{\bullet}$ from the 100mas and the longslit data}

The same was done using only the 100mas dataset together with the
longslit data and the measured PSF. Both $M_\bullet$ and $\Upsilon$
are in agreement between the different quadrants, but slightly smaller
than for the 25mas data
($M_{\bullet}=1.0\dots2.0\times10^8$~M$_{\odot}$,
$\Upsilon=0.65\dots0.7$). Within $2-3\sigma$ errors they still agree
with the 25mas results. Note that the sphere of influence is well
resolved in both cases. 

To figure out whether the uncertainties in the
PSF could raise a bias in $M_{\bullet}$, the same models were
calculated with the 25mas PSF convolved with the kernel from
Fig. \ref{fig:2}c. The results are virtually identical (see
Tab. \ref{tab:results}). Therefore, slight uncertainties in the PSF
shape do not have a noticeable effect on $M_{\bullet}$ or
$\Upsilon$.

As a cross-check the 25mas data, binned $4\times4$ to achieve the
100mas spaxel size and convolved with the kernel, were also modelled
using the convolved 25mas PSF. In this case the confidence intervals are much wider and
the best-fitting values agree within  $1-2\sigma$ with both the
25mas and the 100mas results. 

The difference between the two scales could be related to the
different spatial coverage, although in that case we would expect a
similar result with larger error bars for the 25mas models. An
increase of $\Upsilon$ towards the centre could also be an
explanation. The global $\Upsilon$ would basically not change in that
case, and a larger black-hole mass could mimic the $\Upsilon$
gradient. A triaxial structure could also cause a systematic
difference between the 25mas and the 100mas scale.

Due to the good overall agreement of the different quadrants this
implies that the assumption of axisymmetry is justified and therefore
the four quadrants of each dataset were folded and averaged.  The
results of the averaged data agree with the results of the individual
quadrants.

\subsubsection{$M_{\bullet}$ from the longslit and a combination of the 25mas and the 100mas data}

\begin{figure*}
\begin{minipage}{168mm}
\centering
\includegraphics[width=\linewidth,keepaspectratio,angle=0]{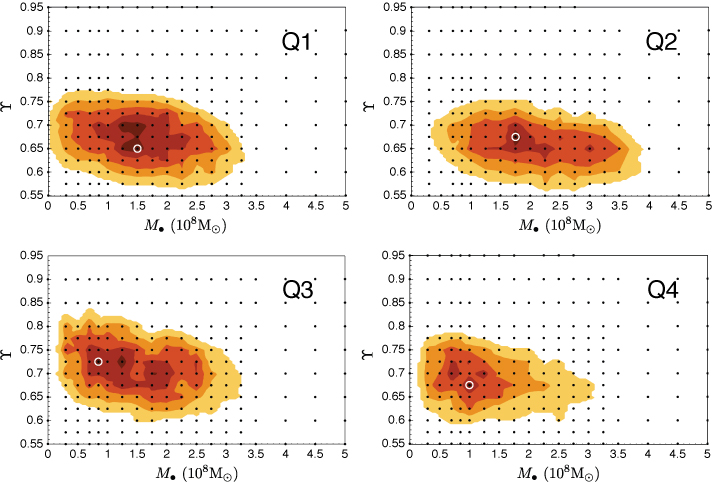}
  \caption{Models calculated for all four quadrants using a
  combination of longslit data out to $31$~arcsec, 100mas and 25mas
  SINFONI data. For each quadrant
  $\Delta\chi_0^2=\chi^2-\chi_{\mathrm{min}}^2$ is plotted as a
  function of the black-hole mass $M_{\bullet}$ and the \emph{Ks} band
  mass-to-light ratio $\Upsilon$. The coloured regions are the
  $1\dots5\sigma$ confidence intervals for two degrees of freedom
  ($\Delta\chi_0^2=2.28$, $6.20$, $12.43$, $19.44$ and $28.65$). Each
  model that was calculated is marked as a black dot, the best-fitting
  model is encircled by a white ring. The $\Delta\chi_0^2$ contours
  are unsmoothed, which sometimes results in disconnected $1\sigma$
  regions due to noise in the models (e.g. in Q3).}\label{fig:15}
\end{minipage}
\end{figure*}

In order to take advantage of the high spatial resolution of the 25mas
dataset and to constrain the orbital distribution of the galaxy
adequately the 25mas data were combined with the 100mas data and this
combined dataset was modelled together with the longslit data. The
spatial region covered already by the 25mas data was not considered in
the 100mas dataset. The results of the dynamical modelling are shown
in Tab. \ref{tab:results} and Fig. \ref{fig:15}, where
$\Delta\chi_0^2$ is plotted as a function of $M_{\bullet}$ and
$\Upsilon$ with error contours for two degrees of freedom. All four
quadrants deliver the same results within at most $3\sigma$ errors
when all data bins are considered (the mean black-hole mass is
$\langle M_{\bullet}\rangle=1.3\times10^{8}$~M$_{\odot}$ with a corresponding
$\mathrm{rms}(M_{\bullet})=0.4\times10^{8}$~M$_{\odot}$ and the mean \emph{Ks} band mass-to-light ratio
is $\langle\Upsilon\rangle=0.68$ with a corresponding $\mathrm{rms}(\Upsilon)=0.03$). When
the inner two radial bins are excluded the results are almost
identical ($\langle M_{\bullet}\rangle=1.2\times10^{8}$~M$_{\odot}$,
$\mathrm{rms}(M_{\bullet})=0.5\times10^{8}$~M$_{\odot}$, $\langle\Upsilon\rangle=0.70$ and
$\mathrm{rms}(\Upsilon)=0.02$). Only the individual $\Delta\chi_{0}^{2}$
contours are wider because of the decreased resolution in the centre.

\begin{figure}
\centering
\includegraphics[width=\linewidth,keepaspectratio]{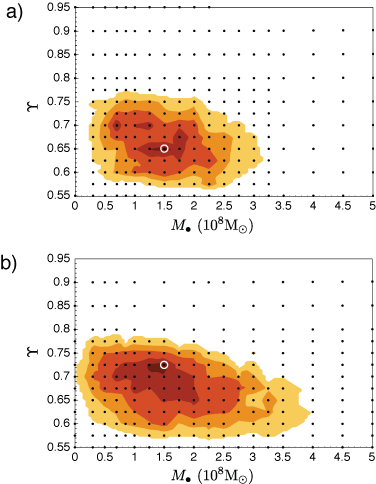}
\caption{Same as Fig. \ref{fig:15} (longslit, 100mas and 25mas data) but for the averaged quadrant with (a) or without (b) the bins of the central two radii.}\label{fig:16}
\end{figure}

\begin{figure}
\centering
\includegraphics[width=\linewidth,keepaspectratio]{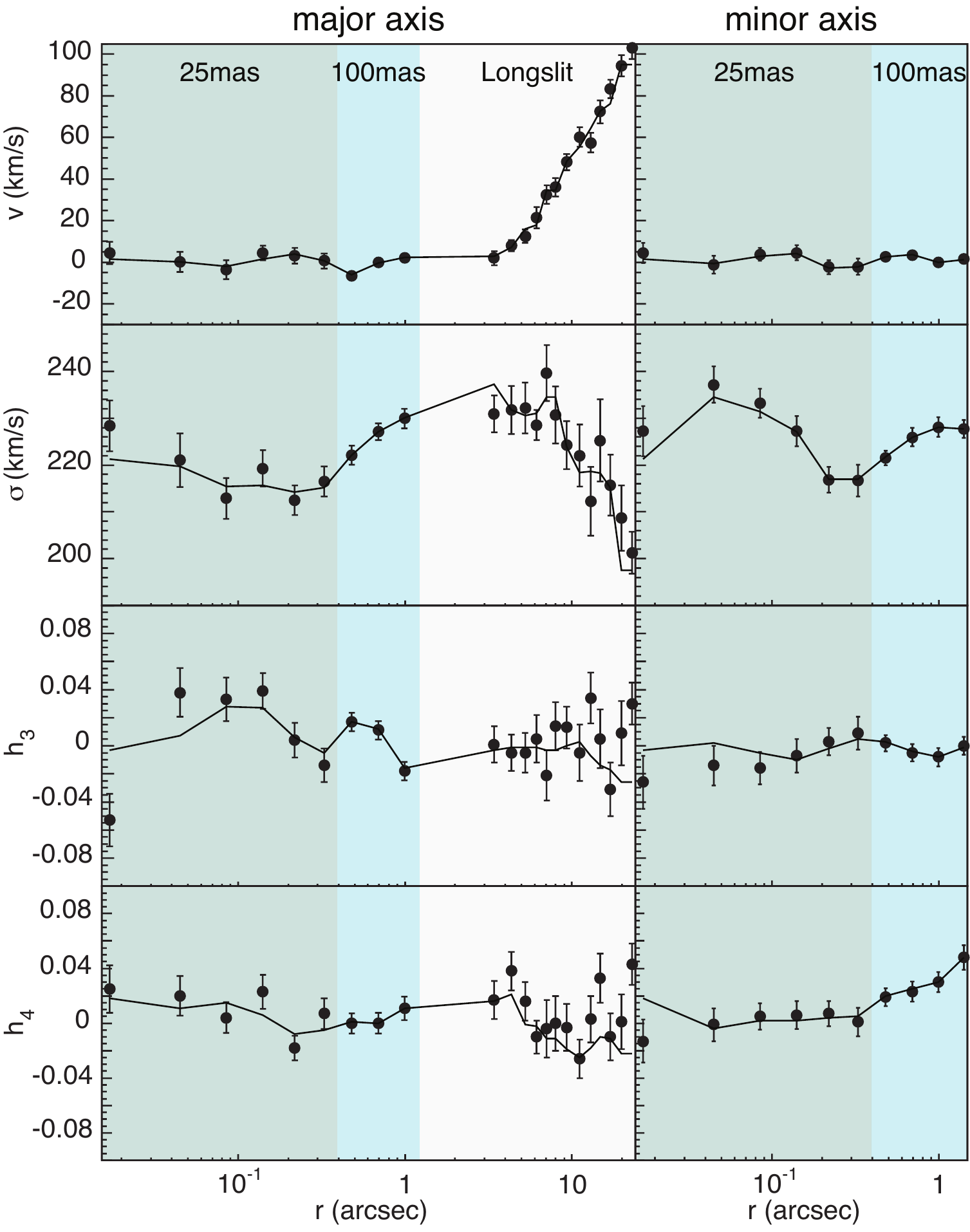}
  \caption{Fit (solid line) of the best model with $M_{\bullet}=1.5\times10^8$~M$_{\odot}$ and $\Upsilon=0.65$ to the major and minor axis kinematics (points). }\label{fig:17}
\end{figure}

\begin{figure*}
\begin{minipage}{168mm}
\centering
\includegraphics[width=\linewidth,keepaspectratio]{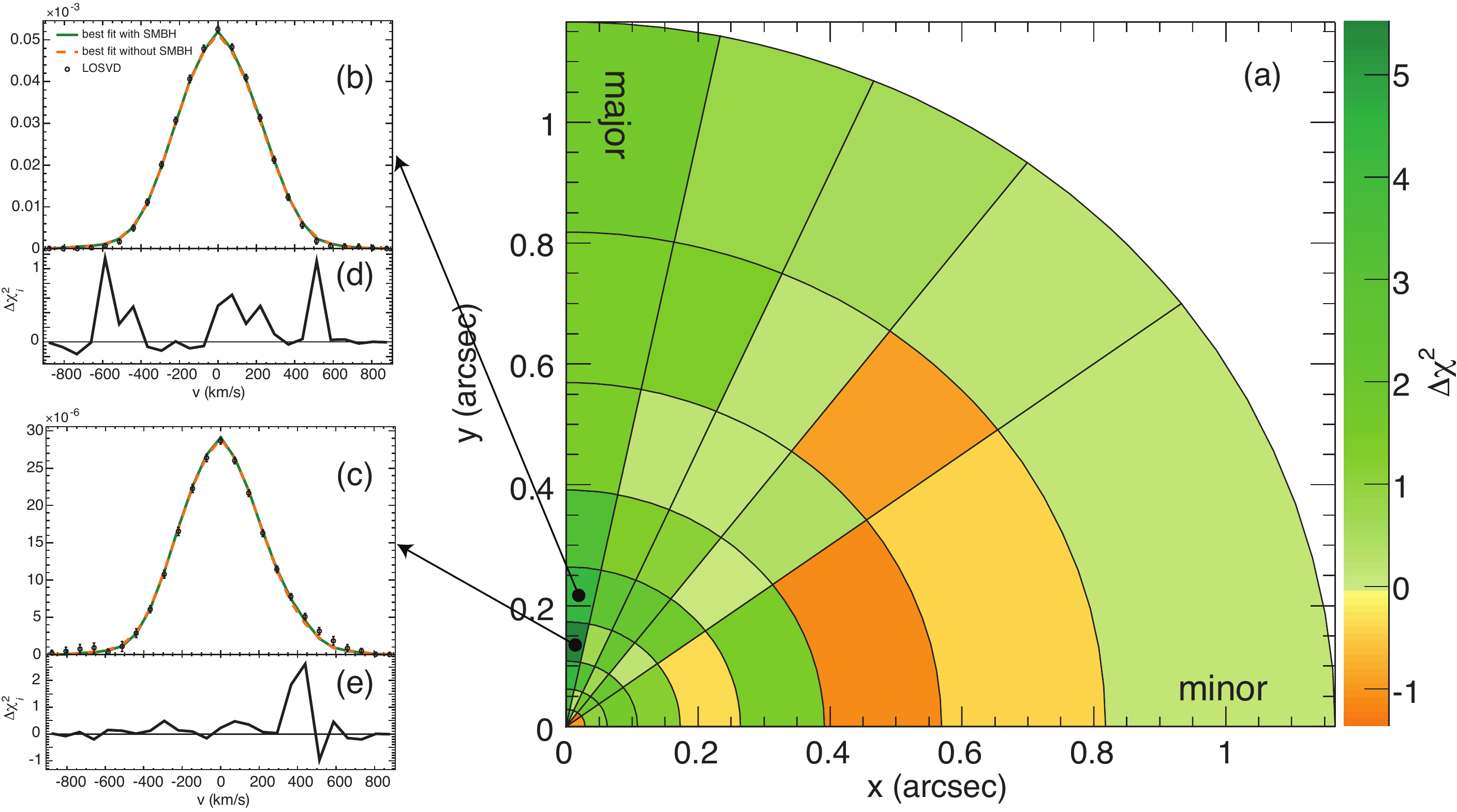}
  \caption{(a) $\chi^2$ difference between the best-fitting model
  without black hole and the best-fitting model with black hole
  ($\Delta\chi^2=\sum_{i}\Delta\chi^2_{i}=\sum_{i=1}^{25}\left(\chi^2_{i,\mathrm{noBH}}-\chi^2_{i,\mathrm{BH}}\right)$
  over all $25$ velocity bins) for all LOSVDs of the averaged quadrant
  (longslit, 100mas and all 25mas SINFONI data). Bins where the model
  with black hole fits the LOSVD better are plotted in green the
  others in orange. (b,c) For the radii with the largest positive
  $\chi^2$ difference in (a) the LOSVD (open circles with error bars,
  normalised as in \citet{Gebhardt-00c}) and both fits (with black
  hole, full green line, and without black hole, dashed orange line)
  are shown with the corresponding $\Delta\chi_i^2$ plotted below
  (d,e).}\label{fig:18}
\end{minipage}
\end{figure*}

As the single quadrants agree well with each other it is legitimate to
just model the LOSVDs folded and averaged over all quadrants. This is shown in
Fig. \ref{fig:16} for both the entire dataset
(Fig. \ref{fig:16}a) and with the central bins excluded
(Fig. \ref{fig:16}b). After the averaging the central velocity
peak disappears, as the LOSVDs of the third and fourth quadrant are
mirrored. The results therefore are in both cases very similar. They
also agree very well with the single quadrants. The black hole mass of
Fornax\,A, derived from modelling the averaged LOSVDs of the combined
25mas and 100mas SINFONI dataset and longslit data, is
$M_{\bullet}=1.5^{+0.75}_{-0.8}\times10^8$~M$_{\odot}$ and the
according mass-to-light ratio is $\Upsilon=0.65_{-0.05}^{+0.075}$
($3\sigma$ errors). The fit of the best-fitting model to $v$,
$\sigma$, $h_{3}$ and $h_{4}$ along the major and the minor axis is
shown in Fig. \ref{fig:17}. Note that the best fit without black hole
would be hardly distinguishable from the shown fit with black hole
(see discussion below and Fig. \ref{fig:18}).  The result does not
change when the inner two radial bins, where the AGN emission distorts
the CO absorption bands are excluded. Only the error bars become
somewhat larger
($M_{\bullet}=1.5^{+1.25}_{-1.2}\times10^8$~M$_{\odot}$ and
$\Upsilon=0.725_{-0.125}^{+0.025}$). Note that the $1\sigma$ errors
derived from a smoothed version of the $\chi^{2}$ profile agree well
with the rms of the four quadrant's solution (see above).

\begin{table*}
 \centering
 \begin{minipage}{140mm}
  \caption{Resulting black-hole masses $M_{\bullet}$ and \emph{Ks} band mass-to-light ratios $\Upsilon$ with the according $3\sigma$ ($\Delta\chi_0^2=12.43$) errors for all modelled datasets and quadrants.}\label{tab:results}
  \begin{tabular}{llllllll}
  \hline
   Dataset & Quadrant & $M_{\bullet}$ [$10^8$~M$_\odot$] & $M_{\bullet,-}$\footnote{Lower $3\sigma$ limit of $M_{\bullet}$ [$10^8$~M$_{\odot}$]} & $M_{\bullet,+}$\footnote{Upper $3\sigma$ limit of $M_{\bullet}$ [$10^8$~M$_{\odot}$]} & $\Upsilon$ & $\Upsilon_{-}$\footnote{Lower $3\sigma$ limit of $\Upsilon$} & $\Upsilon_{+}$\footnote{Upper $3\sigma$ limit of $\Upsilon$}\\
 \hline
 25mas\footnote{SINFONI data with a field of field of view of $0.8\times0.8$~arcsec}+LS\footnote{Longslit data from the Siding Spring $2.3$m telescope \citep{Saglia-02}} & $1$              & 2.0  & 0.3  & 3.5  & 0.725 & 0.625  & 0.775 \\
                      & $2$              & 3.25 & 1.5  & 4.0  & 0.675 & 0.625  & 0.75  \\
                      & $3$              & 2.5  & 0.7  & 3.25 & 0.75  & 0.675  & 0.825 \\
                      & $4$              & 2.0  & 0.3  & 4.0  & 0.775 & 0.675  & 0.825 \\
                      & $1-4$ a\footnote{The LOSVDs of all four quadrants were averaged and then modelled}    & 1.75 & 0.7  & 3.0  & 0.7   & 0.625  & 0.75  \\
                      & $1-4$ a c\footnote{Central two radial bins excluded due to AGN contamination}  & 3.0  & 0.3  & 4.5  & 0.7   & 0.65   & 0.75  
\vspace{.5cm} \\
 100mas\footnote{SINFONI data with a field of view of $3.0\times3.0$~arcsec together with the measured PSF were modelled}+LS     & $1$              & 1.5  & 0.7  & 2.5  & 0.65  & 0.625  & 0.7   \\
                      & $2$              & 1.25 & 0.85 & 2.75 & 0.7   & 0.6    & 0.7   \\
                      & $3$              & 1.0  & 0.3  & 2.0  & 0.675 & 0.65   & 0.75  \\
                      & $4$              & 2.0  & 0.5  & 3.0  & 0.675 & 0.625  & 0.725 \\
                      & $1-4$ a          & 1.0  & 0.3  & 1.75 & 0.65  & 0.6    & 0.725 
\vspace{.5cm}\\
 100mas conv.\footnote{100mas SINFONI data were modelled together with the PSF from Fig. \ref{fig:2}d}+LS  & $1$           & 1.5  & 0.5  & 2.0  & 0.65  & 0.65   & 0.7   \\
                      & $2$              & 1.5  & 0.7  & 2.5  & 0.7   & 0.65   & 0.7   \\
                      & $3$              & 1.0  & 0.3  & 1.5  & 0.7   & 0.65   & 0.75  \\
                      & $4$              & 2.0  & 0.5  & 2.5  & 0.7   & 0.65   & 0.7   
\vspace{.5cm}\\
 25mas conv.\footnote{25mas SINFONI data, binned to the 100mas spaxel scale. Data and PSF were then convolved with the kernel from Fig. \ref{fig:2}c} +LS & $1$              & 2.0  & 0.0  & 4.0  & 0.7   & 0.65   & 0.8   \\
                      & $2$              & 1.5  & 0.3  & 4.0  & 0.7   & 0.65   & 0.8   \\
                      & $3$              & 0.7  & 0.3  & 3.5  & 0.8   & 0.7    & 0.85  \\
                      & $4$              & 1.0  & 0.3  & 2.5  & 0.75  & 0.7    & 0.8   
\vspace{.5cm}\\
 25mas+100mas+LS      & $1$              & 1.5  & 0.3  & 2.75 & 0.65  & 0.625  & 0.725 \\
                      & $2$              & 1.75 & 1.0  & 3.5  & 0.675 & 0.625  & 0.7   \\
                      & $3$              & 0.85 & 0.3  & 2.5  & 0.725 & 0.65   & 0.8   \\
                      & $4$              & 1.0  & 0.5  & 1.75 & 0.675 & 0.625  & 0.75  \\
                      & {\bf 1--4 a} & {\bf 1.5}  & {\bf 0.7}  & {\bf 2.25} & {\bf 0.65}  & {\bf 0.6}    & {\bf 0.725} \\
                      & $1$ c            & 1.5  & 0.3  & 3.0  & 0.7   & 0.625  & 0.75  \\
                      & $2$ c            & 1.75 & 0.5  & 3.0  & 0.675 & 0.625  & 0.725 \\
                      & $3$ c            & 0.85 & 0.3  & 2.5  & 0.725 & 0.675  & 0.8   \\
                      & $4$ c            & 0.7  & 0.3  & 2.0  & 0.7   & 0.65   & 0.75  \\
                      & $1-4$ a c        & 1.5  & 0.3  & 2.75 & 0.725 & 0.6    & 0.75  \\
\hline
\end{tabular}

\medskip
Note. When only one SINFONI dataset was included in the
modelling, the $M_{\bullet}$-$\Upsilon$ parameter space was sampled
more coarsely than for the combined SINFONI dataset, with only about
half the number of models per quadrant. This can sometimes result in
the lower or upper limits being identical to the best-fitting values.
\end{minipage}
\end{table*}

In order to illustrate the significance of the result and where the
influence of the black hole is largest, Fig. \ref{fig:18} shows the
$\chi^2$ differences between the best-fitting model without black hole
and the best-fitting model with black hole for all LOSVDs of the
averaged quadrant, analogous to fig. 7 in \citep{Nowak-07}. As in the
case of NGC\,4486a the largest black hole signature is found within
about $2$ spheres of influence and in particular along the major
axis. For about $85$~per~cent of all bins the model with black hole
produces a fit to the LOSVD better than the model without black
hole. Adding a black hole improves the fit everywhere and not only in
the centre, because orbit-based models have a lot of freedom and will
choose a different orbit distribution if no black hole is assumed. Or
put in another way: a model without a black hole will not just be
worse in the very centre but will be worse over a relatively large
area of the galaxy. For the bins with the largest $\Delta\chi^2$ along
the major axis the LOSVD and the fits with and without black hole are
shown in the left part of Fig. \ref{fig:18} together with the
corresponding $\Delta\chi_i^2$ as a function of line-of-sight
velocity. The largest $\chi^2$ differences appear in the high-velocity
tails of the LOSVDs. The total $\chi^2$ difference, summed over all
LOSVDs, between the best model with black hole and the best model
without black hole is $\Delta\chi^2=54.7$, which corresponds to about
$7.1\sigma$. The total $\chi^2$ values for the models are around
$450$. Together with the number of observables ($60$ LOSVD bins
$\times25$ velocity bins) this gives a reduced $\chi^2$ of
$\approx0.3$. This is a reasonable value as the effective number of
observables is smaller due to the smoothing \citep{Gebhardt-00c}.

For the best-fitting model ($\Upsilon=0.65$) the total stellar mass
within $1$ sphere of influence, where the imprint of the black hole is
strongest, is $M_*=1.11\times10^8$~M$_\odot$. If the additional mass
of $M_\bullet=1.5\times10^8$~M$_\odot$ was entirely composed of stars,
the mass-to-light ratio would increase to
$\Upsilon=1.53$ (corresponding to 
$\Upsilon_{K}=1.08$). This value would be typical for an old stellar
population (around $8$~Gyr for a Salpeter IMF, or $13$~Gyr for a Kroupa
IMF), which has not been found in Fornax\,A
\citep{Goudfrooij-01b,Kuntschner-00}.

\begin{figure*}
\begin{minipage}{168mm}
\centering
\includegraphics[width=\linewidth,keepaspectratio]{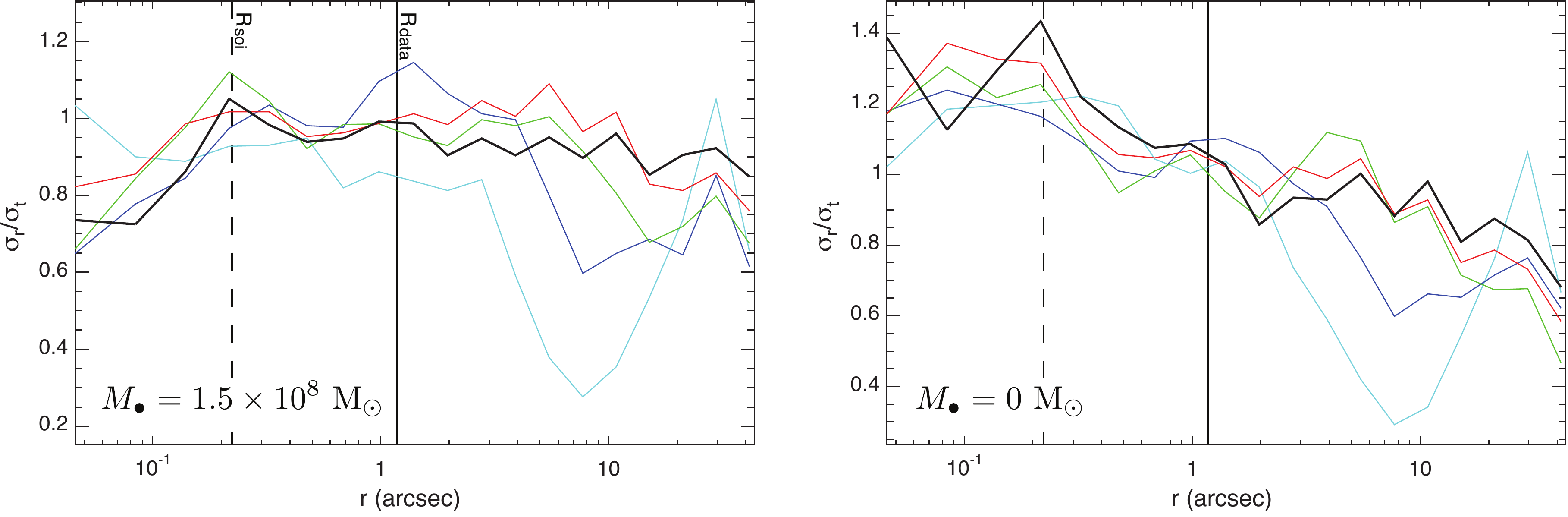}
\caption{Radial over tangential anisotropy as a function of radius for
the best-fitting model with black hole (left) and the best-fitting
model without black hole (right). The values along the major axis are
plotted in black, those along the minor axis in cyan and the other
colours represent the position angles in between. The vertical dashed
line indicates the radius of the sphere of influence and the vertical
solid line marks the radius out to which the SINFONI data used in the
models extend.}\label{fig:19}
\end{minipage}
\end{figure*}

Fig. \ref{fig:19} shows the anisotropy profiles for the best-fitting
model with black hole and the best-fitting model without black
hole. The models with black hole become tangentially anisotropic in
the centre ($r\la0.2$~arcsec), while the models without black hole
show a certain degree of radial anisotropy
($\sigma_r/\sigma_t\approx1.3$ where
$\sigma_t=[(\sigma_\theta^2+\sigma_\phi^2)/2]^{1/2}$).  This behaviour
is not surprising, as a central velocity dispersion increase can be
modelled with either a large black-hole mass or with radial
orbits. However, the no black hole case is a significantly poorer fit
than the best-fitting case with black hole. At large radii the
tangential bias is decreased with an increasing black-hole mass, but
the anisotropy profile in this region is difficult to interpret since
we did not include a dark halo.

\section{Summary \& discussion}
We have obtained near-IR integral field data with three different
spatial resolutions $\geq0.085$~arcsec for the merger remnant
Fornax\,A. Stellar rotation is only detected in the outer parts
($r\ga1.5$~arcsec). The stellar velocity dispersion slightly decreases
from large to smaller radii. In the highest resolution data it shows
two peaks in the centre, separated by a narrow low-$\sigma$
region. The average dispersions are $\sigma=226\pm9$~\kms\ in an 
$8$~arcsec diameter aperture, $\sigma=221\pm4$~\kms\ in a $3.0$~arcsec 
diameter aperture and $\sigma=218\pm8$~\kms\ in an $0.8$~arcsec diameter 
aperture. A low-luminosity AGN is likely to be present and distort the
stellar kinematics in the central $\la0.06$~arcsec, where weak
\caviii\ emission in the region of the second CO bandhead probably induces a
velocity peak. The $\sigma$-drop can be at most partially attributed
to this emission line. Either the AGN continuum emission or a cold
stellar subsystem or both could be the cause.

Near-IR line indices were measured using the high-S/N, intermediate
resolution data to trace stellar populations. All indices show the
same trend, a negative gradient and in the centre an unresolved
depression. The negative gradient of the CO index can be interpreted as a
metallicity gradient. The central drop ($\sim1\mathrm{\AA}$ in CO)
could be, like the $\sigma$ drop, caused by the low-luminosity AGN, a
subsystem with a different stellar population, or a combination of the
two.

Using an axisymmetric Schwarzschild code we find a black hole with the
mass $M_{\bullet}=1.5^{+0.75}_{-0.8}\times10^8$~M$_{\odot}$ ($3\sigma$
C.L. for two degrees of freedom), when the 25mas, the 100mas and the
longslit data are included. The same mass with slightly larger error
bars is found when the innermost bins, where the kinematics is
distorted by the AGN, are excluded from the dynamical modelling. The
assumption of axisymmetry is justified as consistent results were
obtained when modelling single quadrants. The mean black-hole mass
obtained from the four single quadrants is
$\langle M_{\bullet}\rangle=1.3\times10^{8}$~M$_{\odot}$ with
$\mathrm{rms}(M_{\bullet})=0.4\times10^{8}$~M$_\odot$ and the mean \emph{Ks} band mass-to-light ratio
is $\langle\Upsilon\rangle=0.68$ with $\mathrm{rms}(\Upsilon)=0.03$, in agreement with the
$1\sigma$ limit derived from the analysis of the smoothed
$\chi^2$-profile for the averaged data. The PSF of the 100mas data
could not be measured very accurately, but this does not seem to have
noticeable effects. Modelling the 100mas data with the measured or a
reconstructed, noisier PSF (both have about the same FWHM) does not
yield different results.

We find a dynamical \emph{Ks} band mass-to-light ratio of
$\Upsilon=0.65_{-0.05}^{+0.075}$. When the inner two radial bins are
not included it is somewhat larger
($\Upsilon=0.725_{-0.125}^{+0.025}$). This is in agreement with
stellar population models assuming a Salpeter IMF, but only marginally
consistent with a Kroupa IMF. A dark halo plays a significant
role only outside $\sim31$~arcsec, close to the effective radius
($R_\mathrm{e}=36$~arcsec, \citealt{Bedregal-06}). We therefore
included neither a dark halo nor longslit data at radii larger than
$31$~arcsec.

The black-hole mass expected from the $M_{\bullet}$-$\sigma$ relation
\citep{Tremaine-02} is $(2.2\pm0.4)\times10^8$~M$_{\odot}$. Note that
estimating $\sigma_{\mathrm{e}}$, the luminosity-weighted velocity
dispersion within $R_{\mathrm{e}}$, from Fig. \ref{fig:17} gives the
same value as $\sigma$ measured in the $8$~arcsec aperture quoted
above. This mass is consistent with our measurements within the
errors. Fornax\,A is only one of two galaxies that underwent a major
merger a few Gyr ago where a black-hole mass has been measured (the
other one is Cen\,A, \textrm{e.g.}  \citealt{Neumayer-07}). Both seem
to be consistent with the $M_{\bullet}$-$\sigma$ relation, which holds
important implications for the growth of the black hole and its
surrounding bulge. Despite still showing obvious characteristics of
the merging process -- like shells and ripples in the outer envelope
and disordered dust features throughout the entire galaxy -- the black
hole of Fornax\,A has about the mass expected for a normal
elliptical. This suggests that bulges and black holes grow
approximately synchronously in the course of a merger, or that the
growth is regulated during the active phase.

The black-hole mass of Fornax\,A is, however, not in agreement
with the relations between black-hole mass and $K$ band luminosity
$L_K$ or bulge mass $M_{\mathrm{bulge}}$ of
\citet{MarconiHunt-03}. The total 2MASS \emph{Ks} band magnitude
$m_{Ks}=5.587$ corresponds to
$\log(M_{\mathrm{bulge}}/M_{\odot})\approx11.4$ for our
$\Upsilon_{Ks}=0.65$. For this high bulge mass a black-hole mass of
$\sim6\times10^8$~M$_\odot$ would be expected, a factor $\sim4$ larger
than what we measured. Interestingly the situation is similar for
Cen\,A. With the $K$ band magnitude given in \citet{MarconiHunt-03}
and $\Upsilon_K$=0.72 \citep{Silge-05} the bulge mass is
$\log(M_{\mathrm{bulge}}/M_{\odot})\approx11.0$ and a black hole with
$M_\bullet\approx2.2\times10^8$~M$_\odot$ would thus be expected, a
factor $\sim5$ larger than the mass of $4.5\times10^7$~M$_\odot$
obtained by \citet{Neumayer-07}. In this context it would be
interesting to reconsider the relationship between
$M_\bullet/M_{\mathrm{bulge}}$ and galaxy age found by
\citet{Merrifield-00}. Fornax\,A and Cen\,A both have a very small
$M_{\bullet}/M_{\mathrm{bulge}}$ and they would be among the youngest
galaxies in their sample. Although the number of merger galaxies with
measured black-hole masses is fairly small, a correlation between the
time of the last major merger and $M_\bullet$ might start to emerge.

In order to draw more stringent conclusions on the connections between
merging, bulge growth, black hole growth and nuclear activity more
dynamical black-hole mass measurements in merger remnants and in
luminous AGN are necessary. We were able to show that reliable black
hole masses can be derived via stellar dynamical modelling even if the
nuclear source makes dynamical measurements in the centre impossible,
as long as the spatial resolution is high enough that the AGN signature is
well within the sphere of influence.

\section*{Acknowledgments}
We would like to thank the Paranal Observatory Team for support during
the observations. We are very grateful to Harald Kuntschner for
providing us the code to measure near-IR line indices, to Mariya
Lyubenova for helping us implementing the software and numerous
discussions about line indices, and to Yuri Beletsky for providing us
the SOFI images of Fornax\,A. Furthermore we thank Peter Erwin and
Robert Wagner for stimulating discussions. Finally we thank the
referee for his useful comments. This work was supported by the
Cluster of Excellence: ``Origin and Structure of the Universe'' and by
the Priority Programme 1177 of the Deutsche Forschungsgemeinschaft.

\bibliographystyle{mn2e}
\bibliography{bibliography}

\appendix

\section{Dust correction}

The ground-based \emph{Ks} band SOFI and the $H$ band \emph{HST}
NICMOS imaging data were dust-corrected with according $J$ band images
taken with the same instruments using the following method, assuming a
standard extinction law: The relation between reddening and extinction
is

\[
H_0-H_\mathrm{c}=aE_{J-H}
\]
for the \emph{HST} image (analogous for the SOFI image),
where $H_0$ is the observed, and $H_\mathrm{c}$ is the dust-free
image. The absorption coefficient $a$ is defined via the extinctions
in the involved bands: $a=A_{H,K}/(A_{J}-A_{H,K})$.  For the SOFI
image the values $A_J=0.282A_V$ and $A_K=0.112A_V$ from
\citet{Binney-98} were used. The NICMOS image was significantly
overcorrected with $A_J$ and $A_H$ from \citet{Binney-98}. We
found by iteratively changing $a$ that the best correction is obtained
for $a=0.8$.

The observed images provide the fluxes, so with Eq. 1 it follows that

\[
H_\mathrm{c}=H_0-a[(J-H)_\mathrm{0}-(J-H)_c]
\]
and
\[
f_{H_\mathrm{c}}=f_{H_0}\left(\frac{f_{H_0}}{f_{J_0}}\right)^a\left(\frac{f_{J_\mathrm{c}}}{f_{H_\mathrm{c}}}\right)^a\propto\frac{f^{1+a}_{H_0}}{f^a_{J_0}}
\]
under the assumption that
$f_{J_\mathrm{c}}/f_{H_\mathrm{c}}\approx\mathrm{const.}$ (\textrm{i.e.} approximately
no intrinsic stellar population gradient). See also \citet{Beletsky-08}.

\section{Dependence of the reliability of the kinematics on initial and intrinsic parameters}

In order to find the best value for the smoothing parameter $\alpha$
we performed Monte Carlo simulations on a large set of model galaxy
spectra. These were created from stellar template spectra by
convolving them with both Gaussian and non-Gaussian LOSVDs and by
adding different amounts of noise. The results of the simulations
might depend, apart from the S/N, on the spectral resolution, the
LOSVD shape, the size of the velocity bins of the LOSVDs $\delta v$,
the velocity range where the LOSVD is fitted $\Delta v$, the template
itself or other things. Therefore we performed several sets of
simulations using the best-fitting stellar template for each
platescale. The spectral range $\Delta\lambda=2.275\dots2.349~\umu$m
and the number of velocity bins $N=29$ were identical for all
setups. $\delta v$ and $\Delta v$ are given by $N$ and the velocity
dispersion $\sigma$: $\Delta v=2\times4.5\times\sigma$, $\delta
v=\Delta v/N$. We created 100 model galaxy spectra for $\sim35$
S/N-values between $1$ and $170$ and extracted the kinematics of each
spectrum using MPL with $\approx100$ different values for $\alpha$
between $0$ and $500$. We tested the dependence on $\alpha$ for
different S/N values, platescales, templates, $\sigma$, $h_3$ and
$h_4$.

\subsection{Dependence on S/N}
\begin{figure*}
\begin{minipage}{168mm}
\centering
\includegraphics[height=.9\textheight,keepaspectratio]{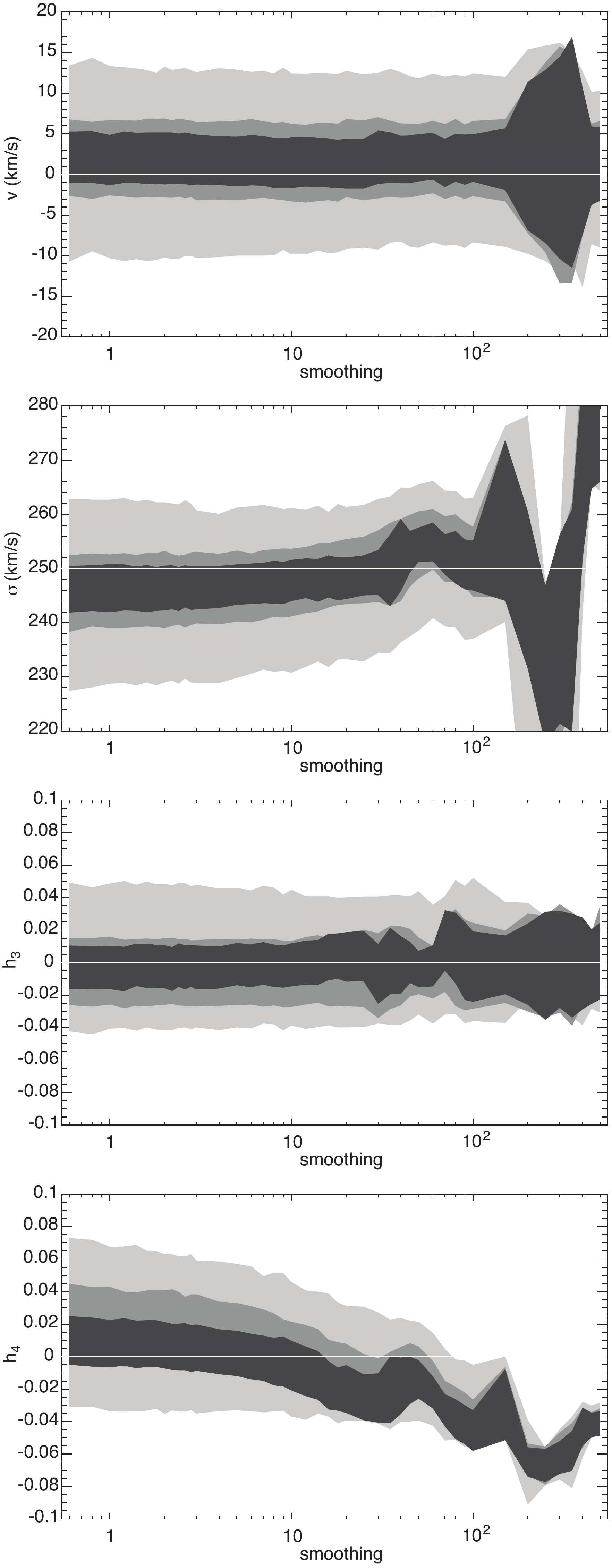}
  \caption{Kinematic parameters for different S/N (light grey: S/N=40,
  medium grey: S/N=90, dark grey: S/N=140). The kinematic template
  star HD12642 was used, observed on the 25mas scale. The spectrum was
  broadened using a LOSVD with $\sigma=250$~\kms\ and
  $h_3=h_4=0$. The reconstructed kinematics is reliable and equal for
  a wide range of $\alpha$-values in the range of S/Ns of our data
  (S/N$\approx70$ for the 250mas and the 25mas and
  $\approx140$ for the 100mas data on average). As expected only the
  error bars increase with decreasing S/N.}\label{fig:B1}
\end{minipage}
\end{figure*}

\begin{figure}
\centering
\includegraphics[width=\linewidth,keepaspectratio]{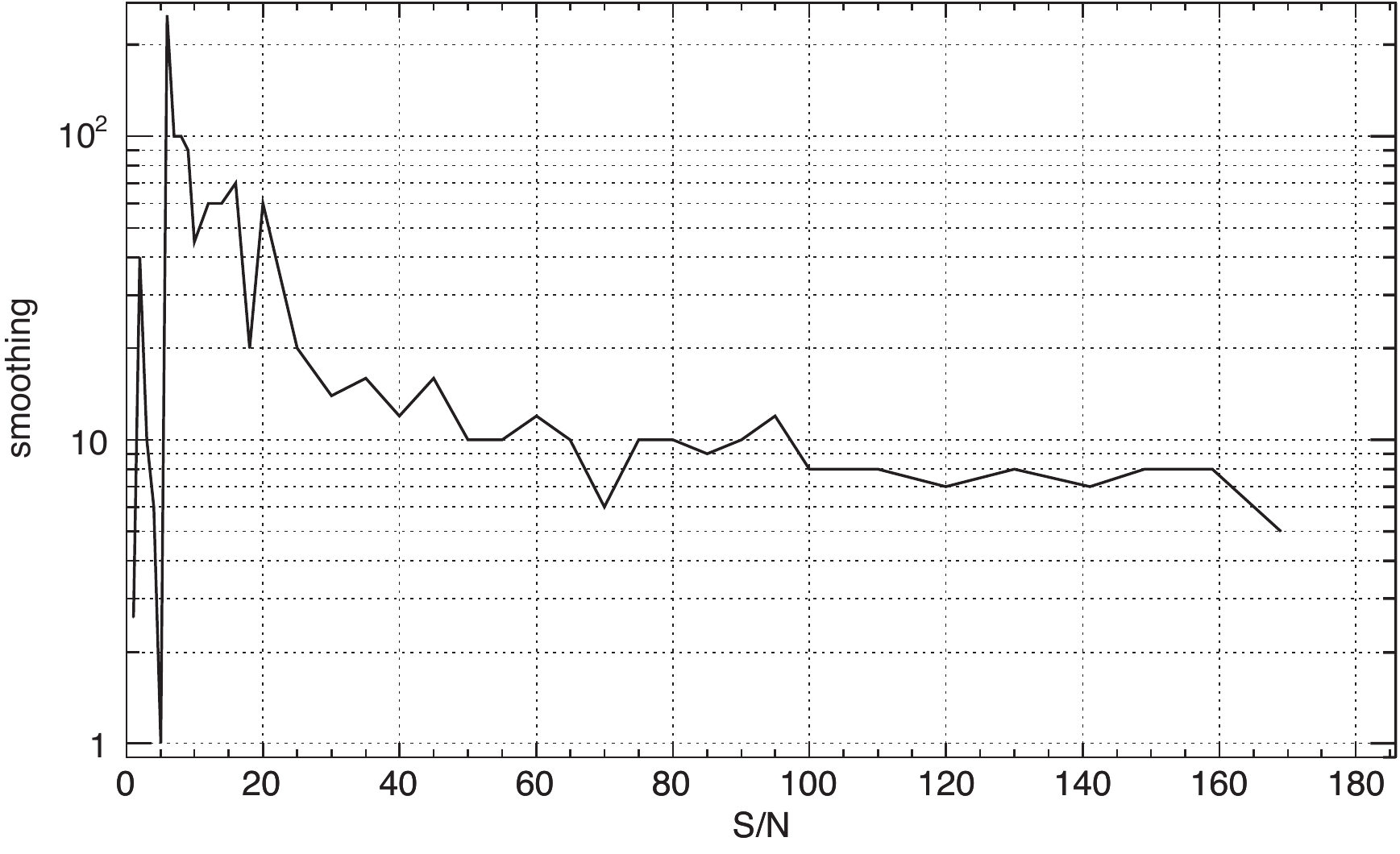}
  \caption{``Optimal'' smoothing parameter as a function of S/N,
  derived from the minimum $\chi^{2}$ between measured and real
  $h_{4}$, the parameter that shows the largest variations with
  $\alpha$. In order to derive reliable kinematics the S/N of the data
  should be $\ga30$.}\label{fig:B2}
\end{figure}

The first set of simulations was done using the template star HD12642,
observed on the 25mas scale. The stellar spectrum was broadened with a
LOSVD with the parameters $\sigma=250$~\kms\ and 
$h_{3}=h_{4}=0$. Different amounts of noise were added
corresponding to a S/N range of $1\dots170$. The kinematic parameters
of the resulting spectra were recovered using the original spectrum as
template star. Fig. \ref{fig:B1}
shows the results for three different S/N values. In general, the
kinematics can be reliably reconstructed for a wide range of smoothing
parameters ($\alpha\approx0.5\dots100$ for $v$ and $h_{3}$,
$\alpha\approx0.5\dots50$ for $\sigma$ and $\alpha\approx0.5\dots20$
for $h_{4}$). Both $\sigma$ and $h_{4}$ change slightly with
$\alpha$. $\sigma$ is biased towards smaller values for small $\alpha$
and then starts to increase. $h_{4}$ is biased towards higher values
for small $\alpha$ and continuously decreases and is biased towards
smaller values for larger $\alpha$. This trend is seen for all
S/N. The only difference is in the error bars, they increase with
decreasing S/N. For small S/N the error bars in $h_{4}$ also increase
with decreasing $\alpha$. As $h_{4}$ is more sensitive to changes in
$\alpha$ than the other parameters, it can be used to determine an
``optimal'' smoothing parameter defined as that $\alpha$ where the
$\chi^2$ difference between measured and real $h_{4}$ is smallest. It
is shown in Fig. \ref{fig:B2} as a function of S/N. For
S/N$\la25$ the curve shows a strong increase and then, for
S/N$<10$, large variations, meaning that it is not possible to
determine reliable kinematics here. For S/N$\ga30$ the curve is
smooth and decreases very slowly, a sign that the kinematics extracted
from such data should be reliable.

\subsection{Dependence on spectral resolution and velocity template}

\begin{figure}
\centering
\includegraphics[height=.9\textheight,keepaspectratio]{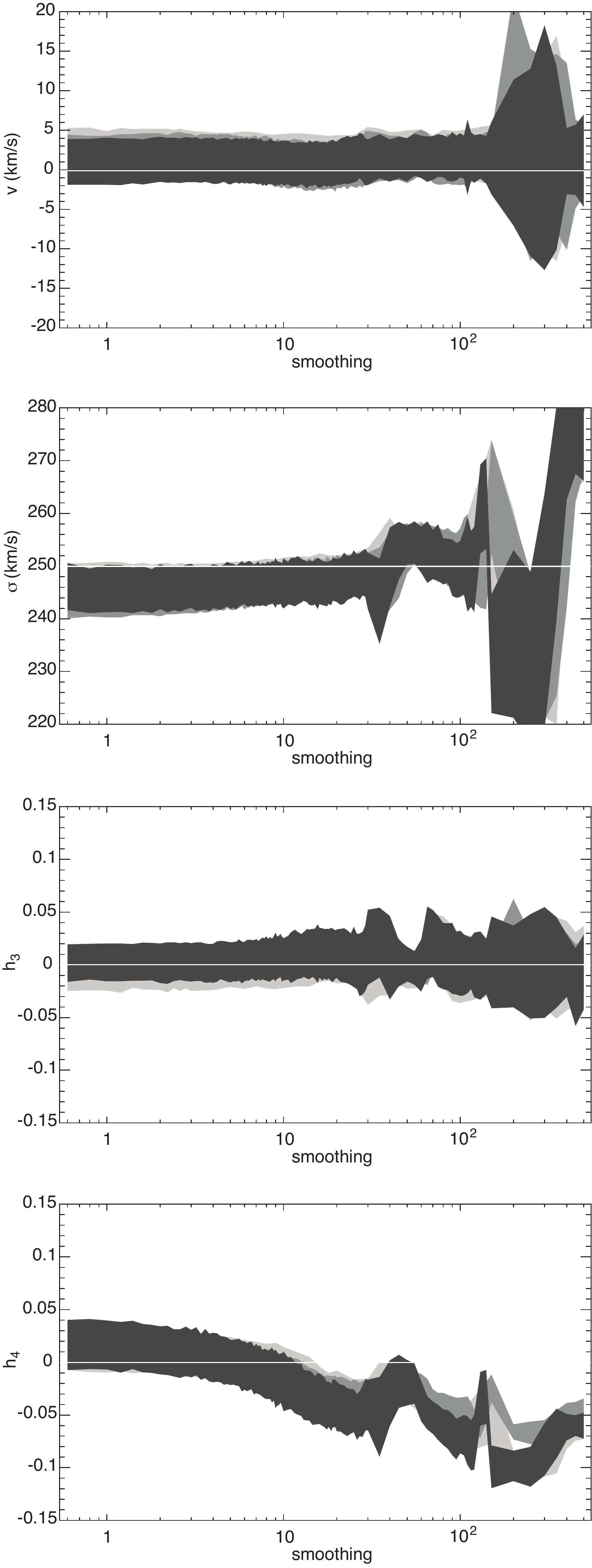}
  \caption{Kinematic parameters for the different platescales (light grey:
  25mas, medium grey: 100mas, dark grey: 250mas). Here the stars HD12642 (25mas) and
  HD181109 (100mas and 250mas) were used, broadened by LOSVDs with
  $v=0$~\kms, $\sigma=250$~\kms\ and $h_{3}=h_{4}=0$.}\label{fig:B3}
\end{figure}

The spectral resolution of the three platescales is slightly
different, $R\approx4500$ for the 250mas scale, $R\approx5000$ for the
100mas scale and $R\approx6000$ for the 25mas scale. This may have an
effect on the derived kinematics or the optimal $\alpha$. The K5/M0III
star HD181109 with a CO equivalent width of $12.9\mathrm{\AA}$ (see
Tab. \ref{tab:templates}) has the largest weight in our library of
velocity templates when fitting to the Fornax\,A spectra. This star
was observed in the 100mas and 250mas scale. In the 25mas scale the
K5III star HD12642 with a similar equivalent width has the largest
weight. For each platescale spectra with $\sigma=250$~\kms,
$h_{3}=h_{4}=0$ and different amounts of noise as above were created
using the according stellar template and the kinematic parameters were
recovered with $\alpha=0.1\dots500$. The results are shown in
Fig. \ref{fig:B3} for S/N$=140$. No significant difference can be
detected between the three platescales. Likewise the optimal smoothing
parameter is equal for all platescales. Note that this set of
simulations has been tailored to the Fornax\,A data and cannot be
generalised to templates with differing CO equivalent widths and for
velocity dispersions close to or below the spectral resolution.

\subsection{Dependence on $\sigma$}

\begin{figure*}
\begin{minipage}{168mm}
\centering
\includegraphics[height=.9\textheight,keepaspectratio]{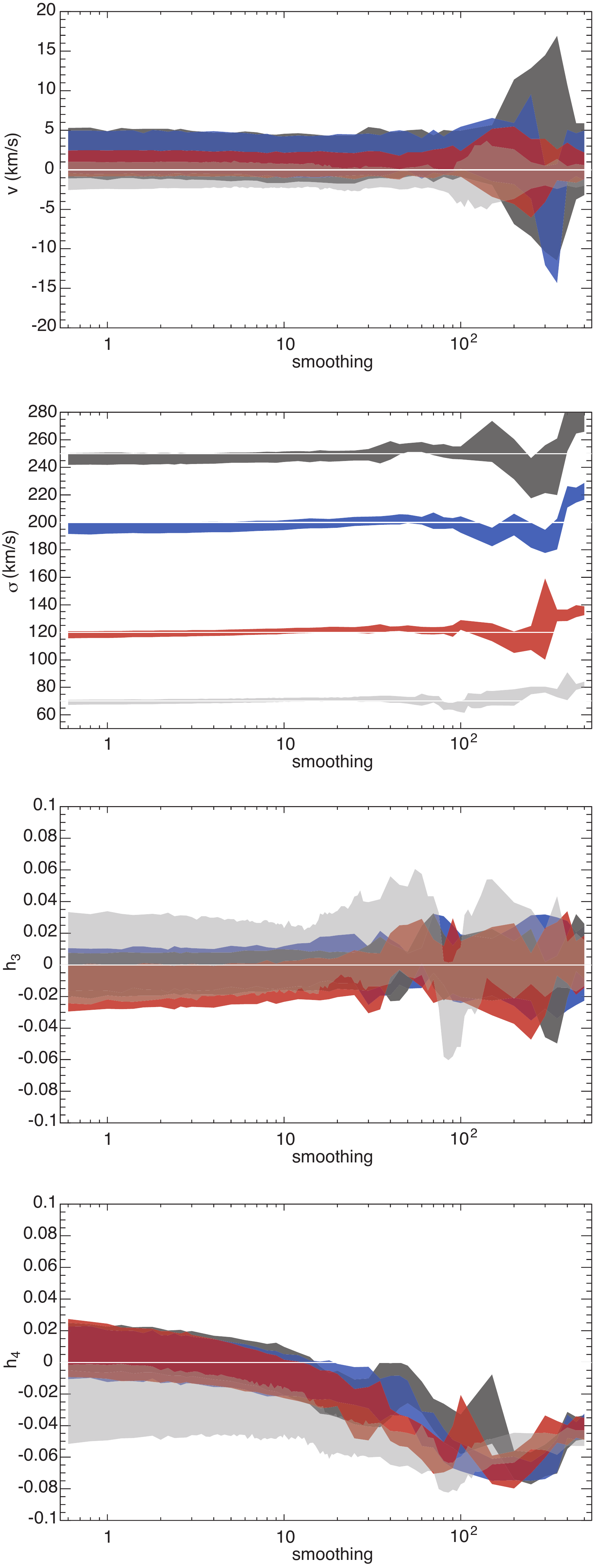}
  \caption{Kinematic parameters for different $\sigma$. Again the star
  HD12642 observed on the 25mas scale was used, broadened by Gaussians
  with $\sigma=250$ (dark grey), $200$ (blue), $120$~\kms\ (red) and $70$~\kms\ (light grey).}\label{fig:B4}
\end{minipage}
\end{figure*}

In the next set of simulations the dependence of the obtained
kinematics on the velocity dispersion of the galaxy $\sigma$ was
tested. The spectrum of HD12642 was broadened with LOSVDs with the
parameters $h_{3}=h_{4}=0$ and $\sigma=70$, $120$, $200$ and
$250$~\kms. The results are shown in Fig. \ref{fig:B4}. The
kinematic parameters for all $\sigma$ values show the same trends with
$\alpha$, but with certain differences. With decreasing $\sigma$ the
LOSVDs become narrower and therefore the scatter in $v$ and $\sigma$
decreases. The error bars of $h_{3}$ and $h_{4}$ are approximately
constant with $\sigma$ except for very low dispersions, where in
addition $h_{4}$ is biased towards negative values. The reason for
that could be that the velocity bin width $\delta v$ is only
$\approx20$~\kms\ for $\sigma=70$~\kms\ (\textrm{i.e.} smaller than
the $30$~\kms~px$^{-1}$ for SINFONI). For low-$\sigma$ galaxies
therefore more simulations are necessary to bypass this effect,
preferentially with $N$ chosen such that $\delta v$ is around
$30$~\kms.

\subsection{Dependence on $h_{3}$ and $h_{4}$}

\begin{figure*}
\begin{minipage}{168mm}
\centering
\includegraphics[width=\linewidth,keepaspectratio]{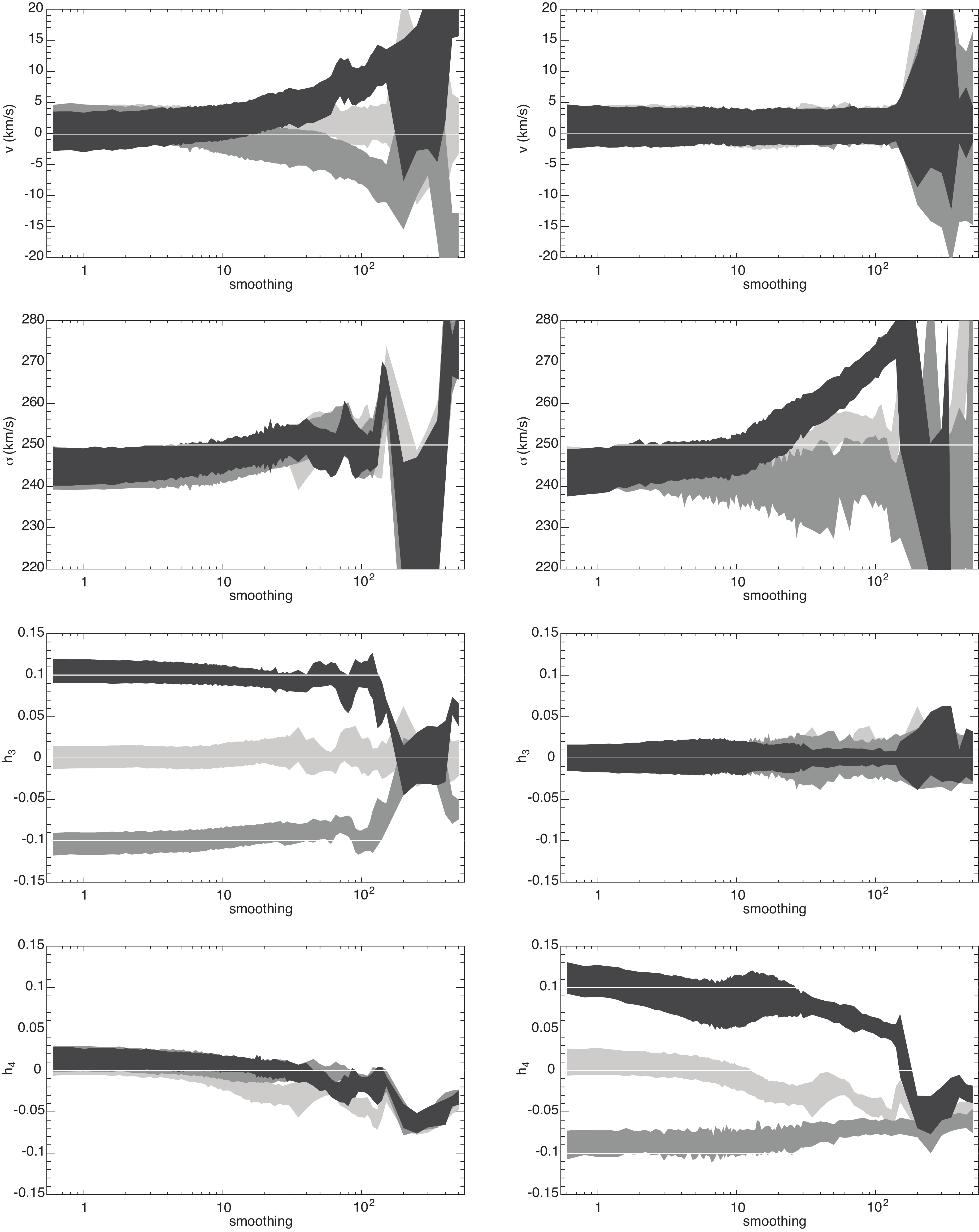}
  \caption{Kinematic parameters for different $h_{3}$ and $h_{4}$.
  Here the star HD181109 observed on the 100mas scale was used. In the
  left column it was broadened by LOSVDs with $\sigma=250$~\kms\ and
  $h_{3}=-0.1$ (medium grey) or $0.1$ (dark grey) and $h_4=0$. In the
  right column the stellar spectrum was broadened by LOSVDs with
  $\sigma=250$~\kms, $h_{3}=0$ and $h_4=-0.1$ (medium grey) or $0.1$
  (dark grey).
  For comparison, the results for $h_{3}=h_{4}=0$ are plotted in
  light grey in both columns. 
}\label{fig:B5}
\end{minipage}
\end{figure*}

With the last set of simulations we tested if the reconstructed
kinematics depends on $h_{3}$ or $h_{4}$. We created model galaxy
spectra where $h_{3}$ and $h_{4}$ have the values $-0.1$ or $+0.1$ and
with $\sigma=250$~\kms. The results are shown in Fig. \ref{fig:B5}
in comparison to the case with $h_{3}=h_{4}=0$. When $h_{3}\neq0$ we
find deviations from the real values with increasing $\alpha$ mainly
in $v$ and $h_{3}$. For simulations with $h_{4}\neq0$ a similar
behaviour is found for $\sigma$ and $h_{4}$. 

\subsection{Results}
Taking all simulations together we conclude that with MPL reliable
kinematics can be obtained from the CO bandheads of the SINFONI data
of Fornax\,A when the chosen $\alpha$ is between $1$ and $10$ and
S/N$\ga30$. 
The S/N of our data is very high (on average $70$ for the 25mas and
the 250mas data and $140$ for the 100mas data). We are using
$\alpha=8$ for the 25mas and 250mas data and $\alpha=6$ for the 100mas
data.  These results cannot be used unrestrictedly for other datasets,
especially not for different absorption lines, much lower $\sigma$
close to or lower than the spectral resolution, or if the CO
equivalent width is significantly different.

\label{lastpage}

\end{document}